\documentclass[sigconf, nonacm]{acmart}

\usepackage[english]{babel}
\usepackage{amsmath}
\usepackage{hyperref}
\usepackage{natbib}
\usepackage{soul}

\usepackage[linesnumbered,vlined,ruled,commentsnumbered]{algorithm2e}
\usepackage[noend]{algpseudocode}
\usepackage{quantumcircuit}
\usepackage{tikz}
\usepackage{subfig}
\usetikzlibrary{positioning}
\usetikzlibrary{shapes}
\usepackage{pgfplots}
\pgfplotsset{compat=1.9, width=10cm}
\usepackage{epstopdf}
\usepackage[]{siunitx}
\usepackage[flushleft]{threeparttable}

\newcommand{\circEqual}[2]{%
  \begin{tabular}{@{}c@{}c@{}c@{}}
    \begin{tabular}{c}
      #1
    \end{tabular}&
    \begin{tabular}{c}
      {\Large=}
    \end{tabular}&
    \begin{tabular}{c}
      #2
    \end{tabular}%
  \end{tabular}%
}

\begin{document}
\title{A Hardware-Aware Heuristic for the Qubit Mapping Problem in the NISQ Era}

\author{Siyuan NIU}
\affiliation{%
  \institution{LIRMM, Univ Montpellier}
  \city{Montpellier}
  \country{France}
}
\email{siyuan.niu@lirmm.fr}

\author{Adrien Suau}
\affiliation{%
  \institution{LIRMM, Univ Montpellier}
  \city{Montpellier}
  \country{France}\\
  \institution{CERFACS}
  \city{Toulouse}
  \country{France}
}
\email{adrien.suau@cerfacs.fr}

\author{Gabriel Staffelbach}
\affiliation{%
  \institution{CERFACS}
  \city{Toulouse}
  \country{France}
}
\email{gabriel.staffelbach@cerfacs.fr}

\author{Aida Todri-Sanial}
\affiliation{%
  \institution{LIRMM, CNRS}
  \city{Montpellier}
  \country{France}
}
\email{aida.todri@lirmm.fr}

\begin{abstract}
Due to several physical limitations in the realisation of quantum hardware, today's quantum computers are qualified as Noisy Intermediate-Scale Quantum (NISQ) hardware. NISQ hardware is characterized by a small number of qubits ($50$ to a few hundred) and noisy operations. Moreover, current realisations of superconducting quantum chips do not have the ideal all-to-all connectivity between qubits but rather at most a nearest-neighbour connectivity. 
All these hardware restrictions add supplementary low-level requirements. They need to be addressed before submitting the quantum circuit to an actual chip. Satisfying these requirements is a tedious task for the programmer. Instead, the task of adapting the quantum circuit to a given hardware is left to the compiler. In this paper, we propose a Hardware-Aware mapping transition algorithm (HA) that takes the calibration data into account with the aim to improve the overall fidelity of the circuit. Evaluation results on IBM quantum hardware show that our HA approach can outperform the state of the art both in terms of the number of additional gates and circuit fidelity.

\end{abstract}

\maketitle

\section{Introduction}
\label{sec:introduction}
In recent years, quantum computing has become a very active field of research. It promises to solve classically intractable computational problems such as integer factorisation~\cite{shor-factorisation}, quantum chemistry~\cite{cao2019quantum}, linear algebra~\cite{1909.03898v1,Harrow2009,1806.01838v1,PhysRevA.101.022322,vqls-bravo-prieto,vqls-huang}, or optimisation~\cite{1808.09266v1,1704.04992v3,qaoa-original}. Along with algorithms, quantum hardware has attracted the attention of several companies such as IBM, Google, Intel, or Rigetti that have demonstrated quantum chips with 53, 72, 49, and 28 qubits respectively. IBM and Rigetti have also given access to a cloud quantum computing service on which anyone can submit quantum circuits to real quantum hardware. The aforementioned quantum hardware can already be qualified as NISQ hardware~\cite{nisq-preskill}. Still, none of them is fault-tolerant as quantum error correction codes (QECC) are in infancy. Nevertheless, it is believed that even a noisy quantum chip with limited qubit-to-qubit connectivity can be used to solve some classically intractable problems, one of the most promising candidates being quantum chemistry~\cite{cao2019quantum}.

From the algorithm perspective, a new paradigm for quantum algorithms has emerged to take into account the limitations of NISQ hardware -- variational algorithms. Examples of variational algorithms include the Variational Quantum Eigensolver (VQE)~\cite{vqe-original}, the Variational Quantum Linear Solver (VQLS)~\cite{vqls-bravo-prieto,vqls-huang}, or the Quantum Approximate Optimisation Algorithm (QAOA)~\cite{qaoa-original}. However, there is a difference between the quantum program written by the programmer and what can be executed on the current quantum hardware. Quantum programs are written as if they were running on ideal quantum hardware without any noise or physical constraints. But real quantum chips are not ideal -- for example for superconducting devices which are targeted in this paper, current two-qubit gates can at best only be applied between two neighbouring qubits. If we want to perform quantum computations, our quantum circuits must obey such connectivity constraints, which means that a modification of the quantum program is necessary to adapt it to the real quantum device. This problem of adapting a quantum program to given hardware connectivity is called the qubit mapping problem and is the focus of this paper.

The qubit mapping problem can be reformulated as two sub-problems. First, to find an initial mapping, i.e.\ a mapping between the "logical qubits" (as a qubit in a quantum circuit) to the "physical qubits" (as a qubit in a quantum chip). Secondly, to determine a mapping transition algorithm to identify the quantum gates to insert in a quantum circuit such that it complies with the targeted quantum hardware topology. Finding the optimal solution for the qubit mapping problem is likely to be an NP-complete problem as noted in~\cite{qubit-allocation-siraichi}.

Two types of methods have been used to solve the qubit mapping problem. The first method is to reformulate it as a mathematically equivalent problem that can then be solved using a specialised solver. Such mathematical formalism can be Integer Linear Programming (ILP)~\cite{bhattacharjee2017depthoptimal,bhattacharjee2019muqut,lao2019mapping,10.1145/3338852.3339829}, Satisfiability Modulo Theory (SMT)~\cite{10.1145/3297858.3304075,murali2019fullstack}, or even Constraint Programming (CP)~\cite{booth2018comparing,Venturelli2019QuantumCC}. However, these mathematical approaches suffer from long runtime and are difficult to scale up. The second method is to use heuristics to modify the quantum circuit, starting from the first quantum gate and transforming the circuit sequentially by making each gate one after the other hardware-compliant.

Most of the previous works~\cite{saeedi2011synthesis,alfailakawi2014lnn,wille2016look,shrivastwa2015fast,kole2016heuristic} using the second method only adapt for nearest-neighbour connectivity and cannot be directly applicable to actual quantum architectures with nonuniform connections. 
Recently, publications\cite{zulehner2018efficient,qubit-allocation-siraichi,li2019tackling,zhou2020quantum,itoko2020optimization,cowtan_et_al:LIPIcs:2019:10397,2019arXiv191200035G,dynamic-lookahead} that are not restricted to a specific architecture have been released. The algorithm presented in~\cite{zulehner2018efficient} uses a heuristic to find the best permutation at each step of the mapping procedure. Instead of representing a quantum circuit as a fixed sequence of layers,~\cite{Guerreschi_2018} introduces the Directed Acyclic Graph (DAG) that takes into account the dependency and commutativity of quantum gates. A major improvement has been shown by~\cite{li2019tackling} which uses a "forward-backward-forward" mapping algorithm. Moreover, the "look-ahead" strategy has been introduced in the heuristic cost function for further optimisation in some existing works, notably~\cite{zulehner2018efficient,li2019tackling,zhou2020quantum,itoko2020optimization,dynamic-lookahead}. For qubit movement, most of these methods only use \texttt{SWAP} gate. A notable exception is~\cite{itoko2020optimization} that considered both \texttt{Bridge} and \texttt{SWAP} gate. Moreover, most of these works aim to minimise the number of inserted gates and do not consider the noise impact on different qubits.

In \cite{tannu2019not,10.1145/3297858.3304075,ash2019qure,finigan2018qubit,bhattacharjee2019muqut}, calibration data is exploited and the applied approach is to insert additional gates between strongly linked qubits, i.e. qubits linked with a low two-qubit gate error rate. However, these works do not consider a holistic view of the problem such as exploring initial mapping or the heuristic function is not efficient enough to select the best candidate of the inserted gate.

In this work, we follow the second type of methods that consist in developing a heuristic to choose the best \texttt{SWAP} to insert based on calibration data. We propose a Hardware-Aware (HA) heuristic mapping transition algorithm to address the drawbacks mentioned above. Our main contributions can be listed as follows. First, we present a mapping transition algorithm that takes into account the hardware topology and the calibration data to improve the overall output state fidelity and reduce the total execution time. Second, to reduce the number of additional gates required to map the quantum circuit to the quantum chip, our algorithm can select between a \texttt{SWAP} or \texttt{Bridge} gate. Finally, we run our HA algorithm on real quantum hardware and compare with various mapping methods from the literature.

\section{State of the art}
Here, we introduce state of the art on quantum hardware devices and their constraints, focusing mainly on IBM quantum devices. Then, we explain the qubit mapping problem. Finally, a small motivational example is shown to illustrate the gist of our algorithm. The notations used in this paper are summarised in~\autoref{tab:title1} (we reference some notations from \cite{li2019tackling}).

\begin {table}[!t]
\caption {Notations used in this paper} \label{tab:title1} 
\begin{center}
\includegraphics{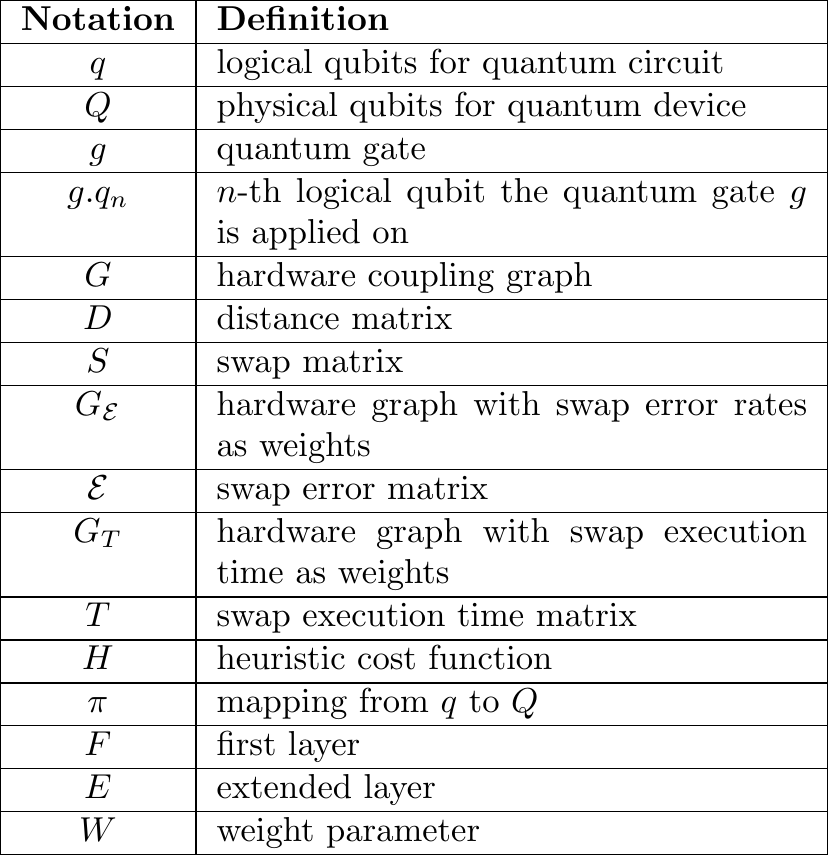}
\end{center}
\end{table}

\subsection{Current state-of-the-art on quantum hardware}\label{sec:current-state-art}
NISQ hardware is characterized in~\cite{nisq-preskill} as quantum hardware having from $50$ to a few hundred noisy qubits on which one can only perform noisy operations. At the time of writing, several companies have already demonstrated quantum chips that can, according to the definition, be qualified as NISQ chips. For example, IBM announced the latest $53$-qubit quantum chip and gave access to the community to execute quantum circuits.
Other companies like Google (72 qubits) or Intel (49 qubits) announced quantum chips that could be qualified as NISQ but did not provide any information about their characteristics. One of the significant challenges of these quantum chips that limits them from solving real-world problems is their level of noise -- even if these chips have enough qubits theoretically to show a quantum speedup, the fidelity of their quantum operations is still too low to obtain any advantage over the classical computer on real-world problems. In this paper, we mainly focus on IBM architectures. Other hardware such as Google's Sycamore~\cite{google-quantum-supremacy} or Rigetti's Aspen-7 is not specially targeted. Still, the proposed algorithm and methods are general enough to be applicable to any quantum chip that use the quantum-gate model of computation and so should be applicable to these hardware.

\begin{figure}
\centering
\includegraphics[width=0.25\textwidth]{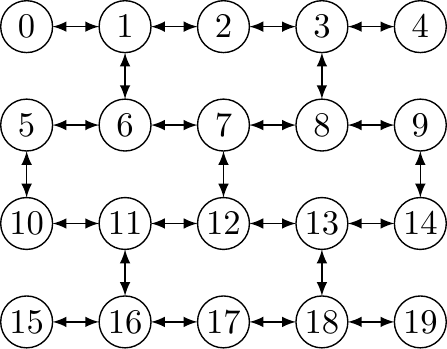}
\caption{ibmq\_almaden topology. Qubits are represented as circles and indexed from $0$ to $19$. A connection between two qubits is represented by an edge between the two qubits.}
\label{fig:ibmq20}
\end{figure}

\begin{table}
\caption{ibmq\_almaden characteristics}
    \label{tab:almaden_caracteristics}
    \centering
    \begin{threeparttable}
    \centering
    \includegraphics{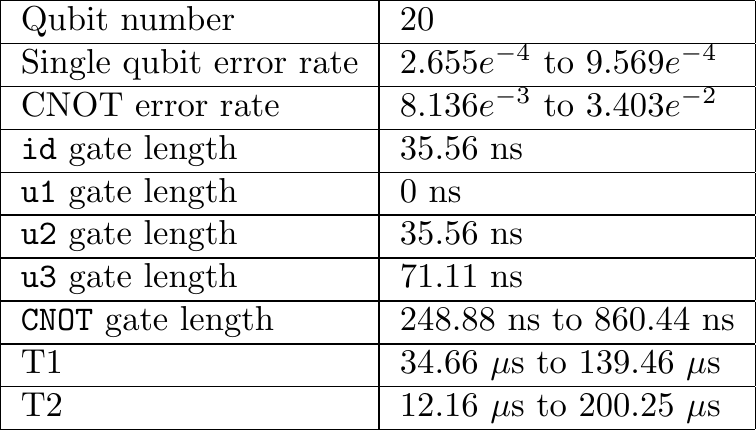}
    \begin{tablenotes}
    \footnotesize
    \item Note that the exact hardware characteristics are not constant and change at each re-calibration of the chip.
    \end{tablenotes}
\end{threeparttable}
\end{table}

Fig.~\ref{fig:ibmq20} shows the topology, also called coupling graph, of IBM Quantum's ibmq\_almaden, a 20-qubit system. Each vertex represents a qubit and the edge represents the coupling interconnect between two qubits. \autoref{tab:almaden_caracteristics} shows the calibration data that are extracted from~\cite{ibmq.hardwareinfos}. It includes \texttt{CNOT} error rates, single qubit error rates, energy relaxation and decoherence characteristic times T1 and T2, and execution time (gate length). The calibration data show that the error of two-qubit gates is one order of magnitude higher than their one-qubit counterparts. This is also the case for gate execution times -- two-qubit gates are approximately an order of magnitude slower than one-qubit gates. For simplicity and because of the relatively low error rates and execution times of one-qubit gates when compared to two-qubit gates, we focus on two-qubit gates in this paper. 

Moreover, it is important to note that all the interconnects between qubits are not equal with respect to \texttt{CNOT} gate error rate or execution time. Taking ibmq\_almaden as an example, the best \texttt{CNOT} gate has an error rate of $4.18$ times lower than the worst \texttt{CNOT} and the maximum execution time is $3.46$ times longer than the minimum one. Therefore, we cannot treat each qubit equally, and we need to consider the interconnect topology between qubits as well as their error rate. 
\texttt{CNOT} gates can be applied in either direction by conjugating with H gates. As we do not consider one-qubit gates in this study, we do not have to consider the connectivity direction.
 
\subsection{Qubit mapping problem}
Following the abstraction first introduced in classical computing decades ago, most of the quantum circuits are described in a generic manner that does not take into account all the physical hardware constraints. Many of the currently existing frameworks for quantum algorithm development encourage this way of development by giving access to a broad set of "primitive" gates. For example, the Qiskit library allows the developer to choose from more than $30$ primitive gates, whereas the IBM quantum chips only provide four physical hardware gates (five if we take the identity gate into account). However, any gate can also be implemented with OpenPulse~\cite{mckay2018qiskit} which is a low level hardware control for users to generate their gates to mitigate errors. Such an abstraction relieves the burden from the developer to adapt the code to a specific hardware and transfer it to the compiler, whose role is to transform an abstracted code into the most efficient hardware code possible. To do so, a compiler for quantum programs should perform several steps summarised in the following paragraphs.

The first step is to decompose the abstracted quantum gates into hardware gates. The hardware gates available strongly depend on the quantum hardware we are compiling for, but are generally comprised of a two-qubit "entangler" gate (controlled-\texttt{X} gate for IBM hardware, fSim gate for Google hardware) and several one-qubit gates (\texttt{u1}, \texttt{u2}, \texttt{u3}, and \texttt{id} gates for IBM hardware). At the end of this step, the quantum circuit has been modified to only contain quantum gates that are directly implemented in the quantum hardware.

But translating all abstract gates to hardware gates is generally not enough to make the quantum circuit executable on the specific hardware -- the hardware topology is rarely respected at the end of this first step and the circuit requires a second step with further modifications. Such modification of the quantum circuit to make it compliant with the hardware topology is often done by inserting \texttt{SWAP} gates before non-executable two-qubit gates. Note that on current hardware, only two-qubit gates are restricted by the hardware topology. 

Finally, once the quantum circuit is executable on the specified quantum hardware, a final third step is performed to optimise the quantum circuit. Depending on the figure of interest, the optimisation can aim at reducing the execution time, gate count, increasing the final state fidelity or even reducing the number of qubits needed.

The qubit mapping problem is defined as the second compilation step that modifies the quantum circuit to contain only two-qubit gates that fit into the hardware topology. But in practice qubit mapping algorithms also try to consider the third step that consists in optimising the generated quantum circuit according to a chosen figure of merit.

\subsection{Motivational example}\label{sec:exec-quant-circ}
\begin{figure*}
    \centering
    \subfloat[Original circuit\label{fig:original_circuit}]{\includegraphics[width=0.25\textwidth,height=3cm]{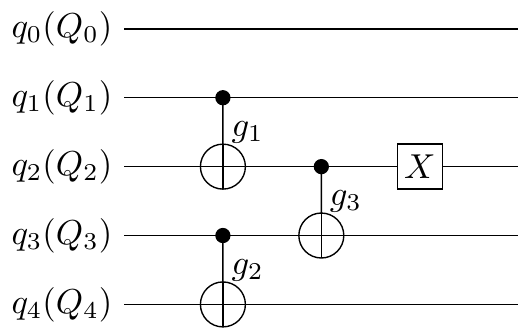}}\hspace{0.1\textwidth}
    \subfloat[ibmq\_valencia \label{fig:ibmq_valencia}]{\includegraphics[width=0.15\textwidth]{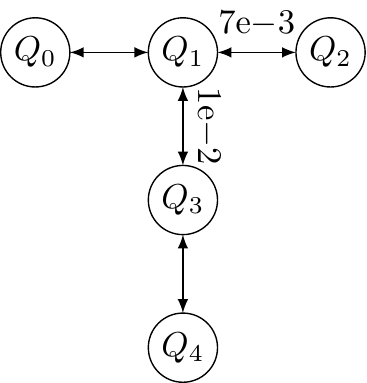}}\hspace{0.1\textwidth}
    \subfloat[Updated circuit\label{fig:updated_circuit}]{\includegraphics[width=0.35\textwidth,height=3cm]{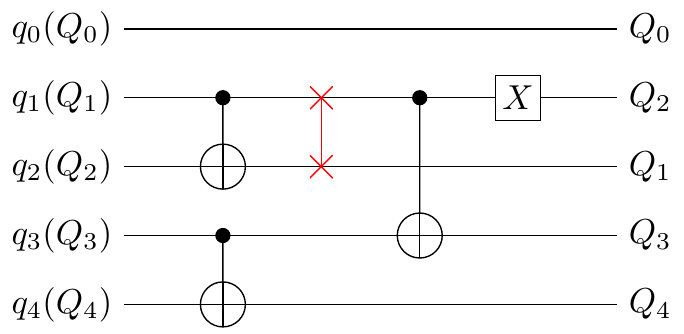}}\qquad
    \caption{A motivational example for the qubit mapping problem.}
    \label{fig:motivation_example}
\end{figure*}
Fig.\autoref{fig:original_circuit} shows a small quantum circuit, which is composed of three \texttt{CNOT} gates and one \texttt{X} gate. It is mapped to a 5-qubit IBM quantum device called ibmq\_valencia, shown in Fig.\autoref{fig:ibmq_valencia}. For simplicity, the initial mapping is allocated linearly as $\{q_0\rightarrow Q_0, q_1 \rightarrow Q_1, q_2 \rightarrow Q_2, q_3 \rightarrow Q_3, q_4 \rightarrow Q_4\}$. Gates $g_1$ and $g_2$ comply with the hardware topology (i.e. coupling constraints) and can be executed directly. However, $g_3$ is applied to two non-connected qubits. Therefore, a movement (i.e. a \texttt{SWAP} gate, shown in Fig.~\ref{fig:swap}) of logical qubits is needed before being able to execute $g_3$ on the hardware connection between $q_2$ and $q_3$. Referring to the coupling graph in Fig.\autoref{fig:ibmq_valencia}, three \texttt{SWAP} gates are possible: $\{q_1,q_2\}$, $\{q_1,q_3\}$ and $\{q_3,q_4\}$. Among these possible \texttt{SWAP}s, two of them change the current mapping between logical and physical qubits in such a way that the \texttt{CNOT} gate between $q_2$ and $q_3$ becomes executable -- swapping of $\{q_1,q_2\}$ and $\{q_1,q_3\}$. 
Translating the logical qubits to their physical counterparts, the \texttt{SWAP}s $\{Q_1, Q_2\}$ and $\{Q_1, Q_3\}$ are our candidates. At this step, most of the state-of-the-art algorithms consider the two possible \texttt{SWAP}s to be equal and will randomly select one. However, if the calibration data is considered, the \texttt{SWAP} between $\{Q_1, Q_2\}$ is less noisy than the other (error rate of the two interconnects is shown in Fig.~\ref{fig:ibmq_valencia}). A \texttt{SWAP} operation consists of three \texttt{CNOT}s and we want to insert a \texttt{SWAP} gate with the least noise. Thus, the \texttt{SWAP} gate between $\{q_1, q_2\}$ is inserted and the final mapping is $\{q_0 \rightarrow Q_0, q_1 \rightarrow Q_2, q_2 \rightarrow Q_1, q_3 \rightarrow Q_3, q_4 \rightarrow Q_4\}$. The updated circuit is shown in Fig.\autoref{fig:updated_circuit}.

\section{Proposed solution}
We are inspired by the SABRE algorithm presented in \cite{li2019tackling}, which is a \texttt{SWAP}-based heuristic algorithm to reduce the number of additional \texttt{CNOT} gates. We propose a Hardware-Aware \texttt{SWAP} and \texttt{Bridge} based heuristic search algorithm.
Compared to SABRE algorithm, which aims at reducing the number of additional gates, we improve the circuit fidelity as well as reduce the number of additional gates by introducing a new distance matrix that takes into account both of the hardware connectivity and the calibration data. Moreover, SABRE only uses \texttt{SWAP} gate when a qubit movement is needed, whereas our algorithm decides between a \texttt{SWAP} and \texttt{Bridge} for qubit movement to further reduce the number of additional gates. Finally, we also develop an initial mapping algorithm called Hardware-aware Simulated Annealing (HSA) in order to evaluate the mapping transition algorithm of different flavours.

The compiler takes as input a quantum program written in the OpenQASM language~\cite{cross2017open} and the calibration data of a specific IBM quantum device. During the compilation process, it considers the hardware constraints such as hardware topology and gate availability. Then, the qubit mapping algorithm is applied. It contains two principal parts -- initial mapping and mapping transition algorithm. In the mapping transition step, some optimisations are done to generate a circuit with a better performance in terms of final state fidelity. The source code is publicly available at \url{https://github.com/peachnuts/HA}.

We start by explaining our HA algorithm in \autoref{sec:ha_descr}. In \autoref{sec:initial_mapping}, we describe the hardware-aware simulated annealing (HSA) method for initial mapping. Finally, \autoref{sec:metrics} presents the metrics used to evaluate our algorithm.

\subsection{Hardware-aware SWAP and Bridge based heuristic search}\label{sec:ha_descr}

The first step of the algorithm is to process the input quantum circuit in order to reformulate it in a more convenient data format. Starting from the input quantum circuit, we can obtain a Directed Acyclic Graph (DAG) circuit which represents the operation dependencies in the quantum circuit without considering the hardware constraints. The DAG is constructed such that quantum gates are represented by the graph nodes and the directed edge $(i, j)$ between nodes $i$ and $j$ represents a dependency from gate $i$ to $j$, i.e. gate $i$ should be executed before $j$.

Once the DAG is constructed, graph nodes (i.e. quantum gates) can be ordered according to the gate dependencies – for example if gate $j$ depends on gate $i$, then gate $i$ will be ordered before gate $j$. One possible ordering that fulfil this property of dependency is the well known topological ordering. Note that depending on the quantum circuit, this ordering might not be unique.

\begin{figure*}
    \centering
    \subfloat[Beginning of the HA mapping algorithm. $g_1$ and $g_2$ do not overlap and are the first gates in the circuit so they are in $F$. The other gates are pushed in $E$.\label{fig:front_layer1}]{\includegraphics[width=0.3\textwidth,height=3cm]{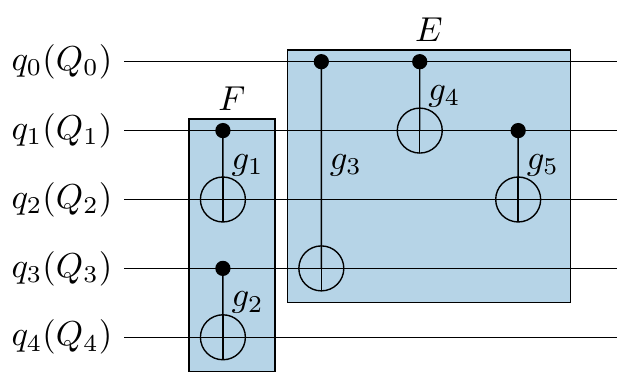}}\hspace{0.03\textwidth}
    \subfloat[$g_1$ and $g_2$ are compliant with the hardware topology. They are executed and removed from $F$. The gate $g_3$ is pushed into $F$ but is not compliant and a SWAP/Bridge should be inserted. $g_4$ overlaps with $g_3$ and cannot be inserted in $F$.\label{fig:front_layer2}]{\includegraphics[width=0.3\textwidth]{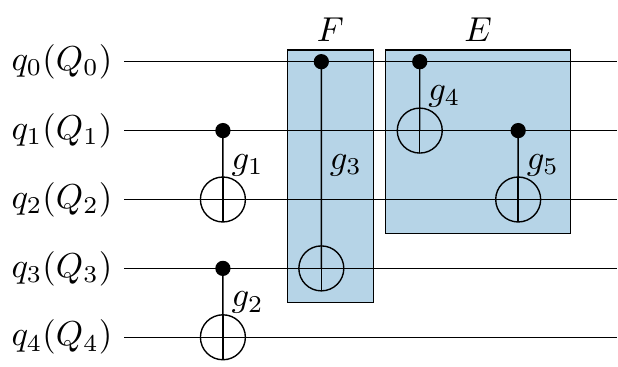}}\hspace{0.03\textwidth}
    \subfloat[After Bridge insertion, $g_3$ is executed and removed from $F$. $g_4$ no longer overlap with a gate in $F$ and is added to the first layer. $g_5$ overlaps with $g_4$ and so should stay in $E$.\label{fig:front_layer3}]{\includegraphics[width=0.3\textwidth,height=3cm]{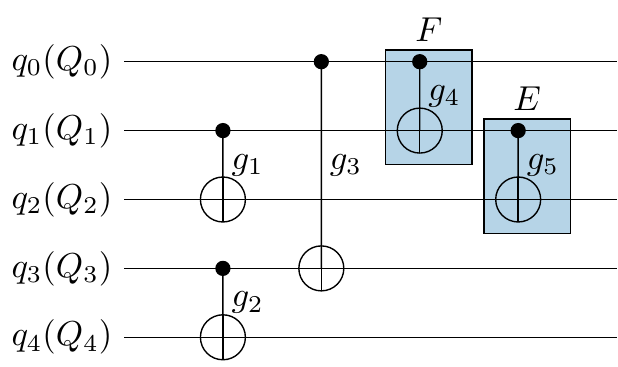}}
    \caption{Evolution of the layers $F$ and $E$ on a simple circuit with a detailed explanation at each step.}
    \label{fig:front_layer}
\end{figure*}

Quantum gates can then be divided into three groups: the executed gates, the executable gates, and to be executed future gates. Executed gates are quantum gates that have already been mapped by the algorithm. Executable gates constitute the first layer, denoted $F$. A gate is considered executable when all the gates it depended on are in the executed gates group. Finally, to be executed future gates are the rest of the gates (not yet executed nor executable). These gates are included in the extended layer, $E$. An illustration of layers $E$ and $F$ is shown in Fig.~\ref{fig:front_layer}.

\subsubsection{Heuristic cost function}
A heuristic cost function $H$ is introduced to estimate the cost of each possible (i.e. executable) swap pairs at a given step of the iterative algorithm. Its objective is to quantify the quality of the possible swap pairs according to the distance considered and to select the best swap pair.

When inserting a \texttt{SWAP} gate, the circuit is divided into two layers: the first layer $F$ and the extended layer $E$. Note that inserting a \texttt{SWAP} gate will not only influence the gates in the first layer $F$ but also the gates in the extended layer $E$. The approach of considering the swap pair's impact on the extended layer is referred as the look-ahead ability. It can contribute to a better selection and depends on the size of the extended layer. 

We devise several metrics that can be used to estimate the cost of a swap pair in HA. We consider three different distance matrices -- swap matrix $S$, swap error matrix $\mathcal{E}$ and swap execution time matrix $T$. Because $S$, $\mathcal{E}$, and $T$ contain entries with incompatible units and different scales, we update $T$ to make it dimensionless and each matrix is normalised. Moreover, we introduce weights (\emph{$\alpha_1$}, \emph{$\alpha_2$}, and \emph{$\alpha_3$} for $S$, $\mathcal{E}$, and $T$, respectively) to allow to choose the importance of each parameter in terms of number of \texttt{SWAP}s, gate error and execution time.

\begin{figure}
    \centering
    \circEqual{%
    \includegraphics{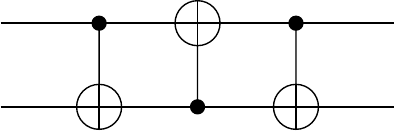}
    }{
    \includegraphics{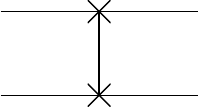}
    }
    \caption{SWAP gate}
    \label{fig:swap}
  \end{figure}

Matrix $S$ is constructed such that the entry $(i, j)$ stores the distance on the real hardware between qubit $i$ to a neighbour of qubit $j$, which is also equal to the minimum number of \texttt{SWAP} gates needed to move qubit $i$ to qubit $j$. The matrix is efficiently constructed by using the Floyd-Warshall algorithm~\cite{floyd1962algorithm}.

Matrix $\mathcal{E}$ stores in its entry $(i, j)$ the minimum error rate attainable to move the qubit $i$ to a neighbour of qubit $j$. The error rate of each possible \texttt{SWAP} is computed based on the calibration data of \texttt{CNOT} gates. The decomposition of a \texttt{SWAP} gate in terms of \texttt{CNOT} gates is shown in Fig.~\ref{fig:swap}. 

The success rate of a \texttt{CNOT} between the physical qubits $Q_i$ and $Q_j$, denoted by $S(Q_i, Q_j)$, is computed from the error rates given in the calibration data. Equation \eqref{eq:2} computes the error rate of a \texttt{SWAP} gate between two connected physical qubits $Q_i$ and $Q_j$ while taking into account that the swap operation is symmetric. The final $\mathcal{E}$ matrix is constructed by using the Floyd-Warshall algorithm on the graph $G_{\mathcal{E}}$ with the computed errors as edge weights.

\begin{multline}
  G_{\mathcal{E}}(Q_i,Q_j) = 1 - S(Q_i,Q_j) \times S(Q_j,Q_i) \\
  \times \max(S(Q_i,Q_j),S(Q_j,Q_i)) \label{eq:2}
\end{multline}

Matrix $T$ is computed, similarly as $S$ and $\mathcal{E}$, with the Floyd-Warshall algorithm applied on graph $G_T$ but by using the \texttt{SWAP} execution time. This execution time is computed with \eqref{eq:3} where $t(Q_i, Q_j)$ is the execution time of the \texttt{CNOT} gate with $Q_i$ as control and $Q_j$ as target, extracted from the calibration data.

\begin{multline}
    G_T(Q_i,Q_j) = t(Q_i,Q_j) 
    + t(Q_j,Q_i) \\
    + \min(t(Q_i,Q_j),t(Q_j,Q_i)) \label{eq:3}
\end{multline}

The summation of the three matrices forms a new matrix called distance matrix $D$ (shown in \eqref{eq:1}). The distance matrix represents the "distance" between each pair of qubits in the quantum chip. Here, the "distance" means the combination of swap distance, overall error rate and execution time of the shortest path.

\begin{equation}
    D = \alpha_1 \times S + \alpha_2 \times \mathcal{E} + \alpha_3 \times T  \label{eq:1}
\end{equation}

Inserting a \texttt{SWAP} gate will have an impact on the current mapping $\pi_c$, changing it to $\pi_{\text{temp}}$. 
We compute the cost of this \texttt{SWAP} on the first layer $F$ with the cost function $H_{basic}$ shown in \eqref{eq:4}. A small score means the \texttt{SWAP} has a little impact on the first layer gates with respect to the overall distance considered. The swap pair with the minimum score is selected as the best candidate.

\begin{equation}
    H_{basic} = \sum_{g \in F} D[\pi_{\text{temp}}(g.q_1)][\pi_{\text{temp}}(g.q_2)] \label{eq:4}
\end{equation}

We also consider the impact of the swap pair on the extended layer $E$. The impact of a \texttt{SWAP} on the first layer is prioritised over its impact on the extended layer. As a result, a weight parameter $W$ is added to the extended layer cost to scale its impact. Moreover, the impacts on the first layer and extended layer are normalised by dividing them with their respective number of gates. The complete heuristic function including the extended layer $E$ with look-ahead ability is shown in \eqref{eq:5}. Even though \eqref{eq:4} and \eqref{eq:5} are similar to equations in \cite{li2019tackling}, it is important to note that the distance matrix $D$ is different.

\begin{multline}
    H = \frac{1}{\vert F\vert}\sum_{g \in F} D[\pi_{\text{temp}}(g.q_1)][\pi_{\text{temp}}(g.q_2)] \\+ W \times \frac{1}{\vert E\vert}\sum_{g \in E} D[\pi_{\text{temp}}(g.q_1)][\pi_{\text{temp}}(g.q_2)] \label{eq:5}
\end{multline}

\begin{figure}
    \centering
    \circEqual{%
    \includegraphics{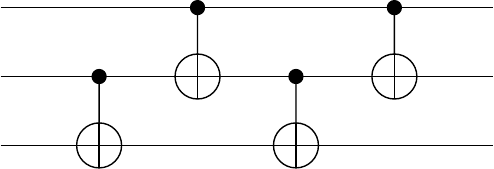}
    }{
    \includegraphics{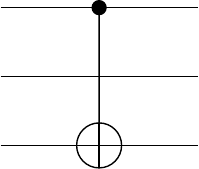}
    }
    \caption{Bridge gate}
    \label{fig:bridge}
\end{figure}

\subsubsection{SWAP gate and Bridge gate}\
Another important metric of the HA algorithm is the heuristic cost function that estimates the usefulness of a \texttt{SWAP}. In some situations, even the best \texttt{SWAP} may have a negative impact on the overall circuit. In that case, inspired by~\cite{itoko2020optimization}, our heuristic function decides to insert a \texttt{Bridge} gate instead of a \texttt{SWAP} gate if the topology allows it. The decomposition of the \texttt{Bridge} gate with four \texttt{CNOT}s is shown in Fig.~\ref{fig:bridge}. The \texttt{Bridge} gate allows executing a \texttt{CNOT} between two qubits that share a common neighbour. Both \texttt{SWAP} and \texttt{Bridge} gate need three supplementary \texttt{CNOT}s. Note that the \texttt{Bridge} gate can only be used to replace a \texttt{CNOT} if the distance between the control and target qubits (i.e. the minimum number of links between the two qubits) is exactly two.

\begin{figure*}
    \centering
    \subfloat[Original circuit\label{fig:original_circuit_b_s}]{\includegraphics[width=0.225\textwidth,height=2.25cm]{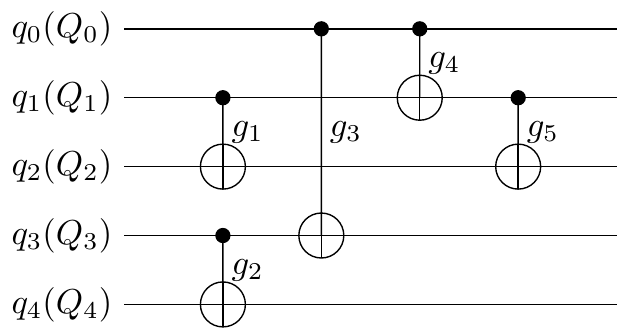}}\qquad
    \subfloat[SWAP gate transformation\label{fig:swap_transfomation}]{\includegraphics[width=0.3\textwidth,height=2.25cm]{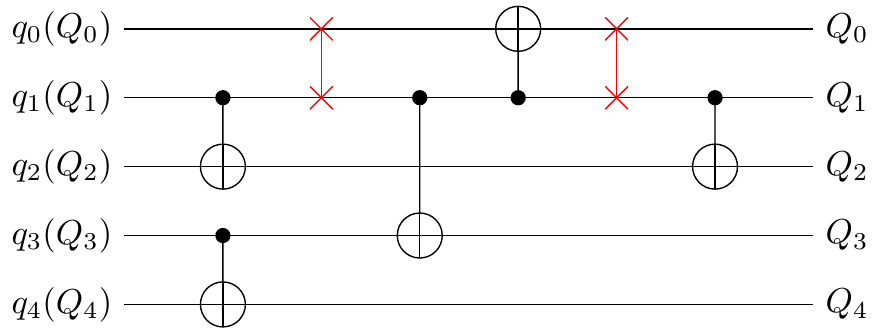}}\qquad
    \subfloat[Bridge gate transformation\label{fig:Bridge_transformation}]{\includegraphics[width=0.35\textwidth,height=2.25cm]{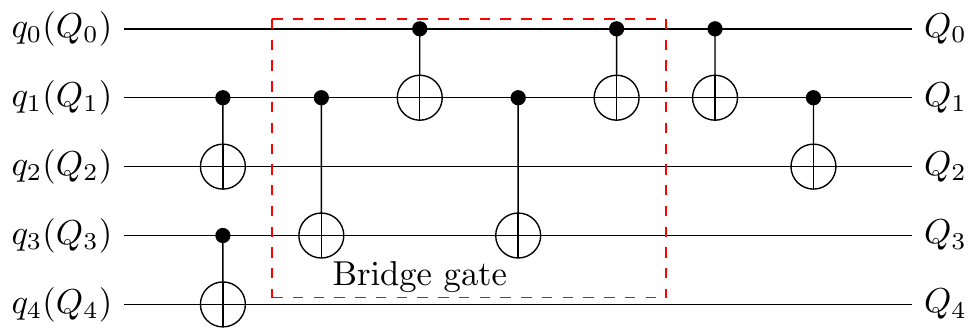}}\qquad
    \caption{An example of a quantum circuit showing the difference between SWAP and Bridge transformation.}
    \label{fig:swap_bridge}
\end{figure*}

Fig.\autoref{fig:original_circuit_b_s} shows an example of quantum circuit that is mapped to ibmq\_valencia with the topology described in Fig.\autoref{fig:ibmq_valencia}. The quantum gates $g_1$ and $g_2$ comply with the topology of the chip, but $g_3$ does not. By evaluating the heuristic cost function $H$, the \texttt{SWAP} between $q_0$ and $q_1$ is selected. But as shown in Fig.\autoref{fig:swap_transfomation}, the chosen \texttt{SWAP} has a negative impact on the extended layer -- gate $g_5$ is no longer executable and another \texttt{SWAP} gate is required to execute it.

Such situations can be solved by using a \texttt{Bridge} gate instead of a \texttt{SWAP} gate as shown in Fig.\autoref{fig:Bridge_transformation}. Since the distance between the control qubit $q_0$ and the target qubit $q_3$ of gate $g_3$ is two, we can insert a \texttt{Bridge} gate instead. Using a \texttt{Bridge} gate allows to execute the \texttt{CNOT} gate $g_5$ without changing the current mapping. Moreover, by using a \texttt{Bridge} gate, we only add three \texttt{CNOT}s to map the entire circuit, instead of six (two times more) if only \texttt{SWAP} gates were used. 

Once the cost $H$ of each swap pair is computed, the heuristic will try to choose the best option between inserting a \texttt{SWAP} or \texttt{Bridge} gate. To do so, it considers two mappings: $\pi_c$, the mapping used before selecting the best swap pair and also the mapping obtained after inserting a \texttt{Bridge} gate, and $\pi_{\text{temp}}$, the new mapping that would be obtained after inserting the best \texttt{SWAP} gate.
The overall effect of the \texttt{SWAP} gate on the extended layer $E$ is computed according to \eqref{eq:6}. If the effect of the best \texttt{SWAP} gate is negative, this means that the considered swap pair has an overall negative impact on the extended layer $E$. In this case, we consider that it is better to keep the current mapping so, if the hardware topology permits it, a \texttt{Bridge} gate is inserted instead of a \texttt{SWAP} gate.

\begin{multline}
    \text{Effect} = \sum_{g \in E} D[\pi_c(g.q_1)][\pi_c(g.q_2)] \\
    - D[\pi_{\text{temp}}(g.q_1)][\pi_{\text{temp}}(g.q_2)] \label{eq:6}
\end{multline}

\subsubsection{HA Algorithm}

The mapping transition algorithm will go through each quantum gate sequentially and mark the directly executable gates as executed. If no more gates can be marked as \emph{executed}, this means that either the quantum circuit is fully mapped or all the gates in the first layer do not comply with the hardware topology. In the first case, the mapping algorithm can be stopped and the mapped quantum circuit returned. In the second case, the algorithm calls a heuristic function to choose the best \texttt{SWAP} or \texttt{Bridge} gate to insert in order to make some of the gates in the first layer executable. The algorithm then iterates, until the quantum circuit is fully mapped.

Algorithm~\ref{algo3} shows the pseudo-code of the HA heuristic method. Note that the most recent calibration data should be retrieved (i.e. through the IBM Quantum Experience) before each usage of the HA algorithm to ensure that the algorithm has access to the most accurate and up-to-date information possible. 
%

The heuristic method to insert a \texttt{SWAP} or \texttt{Bridge} when no gate in the first layer $F$ is executable can be described as follows.
First, a list of all the candidate \texttt{SWAP} gates, \texttt{swap\_candidate\_list}, is constructed based on the quantum gates in the first layer $F$ and the hardware coupling graph $G$.  Then, for each \texttt{SWAP} candidate a temporary mapping $\pi_{temp}$ is computed with the \texttt{Map\_Update} function. 
The final cost of the candidate \texttt{SWAP} is computed following \eqref{eq:5}. The \texttt{SWAP} with the minimum score is selected and called $swap_{min}$.

The last step is to choose between a \texttt{SWAP} gate or a \texttt{Bridge} gate. A \texttt{SWAP} gate can always be used, whereas a \texttt{Bridge} gate can only be inserted if a gate in the first layer $F$ becomes executable from the mapping obtained after applying the $swap_{min}$ gate. If a \texttt{Bridge} gate is not insertable, then the algorithm has no choice but to insert a \texttt{SWAP} gate. Else, the algorithm decides the gate (\texttt{SWAP} or \texttt{Bridge}) to insert based on the effect of the \texttt{SWAP} gate on the extended layer computed with \eqref{eq:6}. If adding a \texttt{SWAP} gate has a negative impact on the extended layer, then a \texttt{Bridge} gate (which does not change the current mapping) is inserted.  Otherwise, if adding a \texttt{SWAP} gate has a positive effect on the extended layer, then the algorithm inserts a \texttt{SWAP} gate.

\begin{algorithm}
\SetAlgoLined
\caption{Heuristic algorithm for selecting additional gate candidate} \label{algo3}
\SetKwInOut{Input}{input}\SetKwInOut{Output}{output}
\SetKwFunction{FindSwapPairs}{FindSwapPairs}
\SetKwFunction{MapUpdate}{Map\_Update}
\SetKwFunction{Append}{append}
\SetKwData{score}{score} 
\SetKwData{cost}{$H$}
\SetKwData{firstCost}{$H_{\text{basic}}$}
\SetKwData{secondCost}{$H_{\text{extended}}$}
\SetKwData{effect}{effect}
\SetKwData{effectCost}{effect\_cost}
\SetKwData{swapCandidateList}{swap\_candidate\_list}
\SetKwData{swap}{swap}
\SetKwData{CircuitDAG}{$DAG$}
\SetKwData{CouplingGraph}{$G$}
\SetKwData{CurrentMapping}{$\pi_c$}
\SetKwData{DistanceMatrix}{$D$}
\SetKwData{SwapMatrix}{$S$}
\SetKwData{FirstLayer}{$F$}
\SetKwData{ExtendedLayer}{$E$}
\SetKwData{newMapping}{$\pi_n$}
\SetKwData{gateInsert}{$g_{\text{add}}$}
\SetKwData{tempMap}{$\pi_{\text{temp}}$}
\SetKwData{gate}{gate}
\SetKwData{swapMinScore}{$\text{swap}_{\text{min}}$}
\SetKwData{gateRelatedSwap}{$g_s$}
\SetKwData{Bridge}{$g_B$}
\SetKwData{Weight}{$W$}
\Input{Circuit \CircuitDAG, Coupling graph \CouplingGraph, Current mapping \CurrentMapping, Distance matrix \DistanceMatrix, Swap matrix \SwapMatrix, First layer \FirstLayer, Extended layer \ExtendedLayer, Weight parameter \Weight}
\Output{New mapping \newMapping, Inserted gate \gateInsert}
\BlankLine
\Begin{
Set \score to empty list\;
Set \effect to empty list\;
\swapCandidateList $\leftarrow$ \FindSwapPairs{\FirstLayer, \CouplingGraph}\;
\For{\swap $\in$ \swapCandidateList}{
    \tempMap$\leftarrow$\MapUpdate(\swap)\;
    \firstCost $\leftarrow$ 0\;
    \For {\gate $\in$ \FirstLayer} {
        \firstCost $\leftarrow$ \firstCost $+$ \DistanceMatrix{\gate, \tempMap} \;
    }
    \secondCost $\leftarrow$ 0\;
    \For {\gate $\in$ \ExtendedLayer} {
        \secondCost $\leftarrow$ \secondCost $+$ \DistanceMatrix{\gate, \tempMap} \;
        \effectCost $\leftarrow$ \effectCost $+$ \DistanceMatrix{\gate, \CurrentMapping} $-$ \DistanceMatrix{\gate, \tempMap}\;
    }
    \cost $\leftarrow \frac{1}{\vert F \vert} \text{\firstCost} + \frac{\text{\Weight}}{\vert E \vert} \text{\secondCost}$ \;
    \score.\Append{\cost}\;
    \effect.\Append{\effectCost}\;
}
\emph{Find the swap with minimum score: \swapMinScore}\;
\emph{Find the gate in \FirstLayer that become executable by applying \swapMinScore: \gateRelatedSwap}\;
\eIf{$\text{\effect}\left[\text{\swapMinScore}\right] < 0$   \textbf{and} \SwapMatrix{\gateRelatedSwap,\CurrentMapping} $= 2$} {
    \newMapping $\leftarrow$ \CurrentMapping \;
    \gateInsert $\leftarrow$ \Bridge\;
} {
    \newMapping $\leftarrow$ \MapUpdate(\swapMinScore)\;
    \gateInsert $\leftarrow$ \swapMinScore\;
}
\Return \newMapping, \gateInsert \;
}
\end{algorithm}

\subsubsection{Runtime analysis}

The HA algorithm outperforms SABRE algorithm thanks to several modifications while not changing its asymptotic complexity. The mapping procedure is separated into two steps: an initialisation step that is independent of the mapped quantum circuit and a mapping step.

The initialisation step computes the distance matrix that is used afterwards in the mapping step. In our algorithm, the distance matrix is computed according to \eqref{eq:4}. Each of $S$, $\mathcal{E}$ and $T$ constituting the distance matrix $D$ requires to use the Floyd-Warshall algorithm once on the hardware graph. This means that we need to perform three calls to an algorithm of $O(n^3)$ complexity, $n$ being the number of qubits of the targeted quantum chip. Moreover, the weights used by the Floyd-Warshall algorithm for the matrices $\mathcal{E}$ and $T$ should be retrieved online with Qiskit API. This retrieval is an operation that theoretically takes $O(n^2)$ time in the worst case as we need to retrieve \texttt{CNOT} error rates and execution time for each link. Note that the current quantum chips only have $O(n)$ links and so the asymptotic complexity of this step is $O(n)$. Overall, the initialisation step is dominated by the cost of applying the Floyd-Warshall algorithm, that takes $O(n^3)$ time.

After the initialisation step, the actual mapping procedure is applied. Let $n$ be the number of qubits, $g$ the number of \texttt{CNOT} gates in the mapped quantum circuit and $d$ the diameter of the chip, i.e. the minimum \texttt{SWAP} distance between the two farthest qubits on the quantum chip. In the worst case, all the \texttt{CNOT} gates should be mapped because none of them comply with the hardware topology. Moreover, all the \texttt{CNOT} gates might need up to $d$ \texttt{SWAP}s in order to become executable. Finally, for each \texttt{SWAP} insertion we need to execute the heuristic cost function. This function will need to explore at most $n^2$ links (in the case of an all-to-all connected chip, this number improves to $O(n)$ on practical quantum chips with a nearest-neighbour connectivity), where exploring one link might take a time of $O(g)$ if all the \texttt{CNOT} gates are included in either $F$ or $E$. In summary, the mapping step takes $O(g^2dn^2)$ time in the worst case, which can be improved to $O(gn^{2.5})$ under reasonable assumptions (nearest-neighbour chip connectivity, i.e. $d \in O(\sqrt{n})$, and an extended layer $E$ with at most $O(n)$ \texttt{CNOT} gates).

It is important to note that the initialisation step only needs to be repeated when the calibration data change but that requires to recover data from the Internet which can be a slow operation (in the order of several seconds).

\subsection{Initial mapping}\label{sec:initial_mapping}

Heuristic-based mapping transition algorithms rely crucially on a good initial mapping to achieve the best results. A well-known algorithm when trying to approximate the global minimum of a scalar function with a discrete search space is simulated annealing. Simulated annealing is a meta-heuristic designed to explore the search space by randomly selecting neighbours of the current state, evaluating them with the provided cost function and evolving in such a way that the algorithm will not be trapped into local minimums. The simulated annealing algorithm is depicted in Algorithm \ref{alg:annealing}.

\begin{algorithm}
\caption{Simulated annealing} \label{alg:annealing}
\SetAlgoLined
\SetKwInOut{Input}{input}\SetKwInOut{Output}{output}
\SetKwFunction{FindSwapPairs}{FindSwapPairs}
\SetKwFunction{costFunction}{C}
\SetKwFunction{getNeighbour}{get\_neighbour}
\SetKwFunction{rand}{rand}

\SetKwData{initialMapping}{$\pi_0$} 
\SetKwData{initialT}{$T_{\text{init}}$}
\SetKwData{finalT}{$T_f$}
\SetKwData{temp}{$T$}
\SetKwData{cost}{cost}
\SetKwData{costOpt}{$\text{cost}_{\text{opt}}$}
\SetKwData{finalMapping}{$\pi_{\text{opt}}$}
\SetKwData{mapping}{$\pi$}
\SetKwData{Tconstant}{$\Delta$}
\SetKwData{neighbour}{$\pi_{\text{neighbour}}$}
\SetKwData{neighbourCost}{$\text{cost}_{\text{neighbour}}$}

\Input{Initial mapping \initialMapping, Cost function \costFunction, Neighbour computation function \getNeighbour, Initial temperature \initialT, Final temperature \finalT, Temperature evolution constant \Tconstant }
\Output{Best initial mapping found \finalMapping}
\BlankLine
\Begin{
\mapping $\leftarrow$ \initialMapping\;
\finalMapping $\leftarrow$ \initialMapping\;
\temp $\leftarrow$ \initialT\;
\cost $\leftarrow$ \costFunction{\mapping}\;
\costOpt $\leftarrow$ \cost\;

\While{\temp $\geqslant$ \finalT}{
    \neighbour $\leftarrow$ \getNeighbour{\mapping}\;
    \neighbourCost $\leftarrow$ \costFunction{\neighbour}\;
    \If{\neighbourCost $<$ \costOpt} {
        \costOpt $\leftarrow$ \neighbourCost \;
        \finalMapping $\leftarrow$ \neighbour \;
    }
    
    \eIf{\neighbourCost $<$ \cost}{
        \cost $\leftarrow$ \neighbourCost \;
        \mapping $\leftarrow$ \neighbour \;
    }{
        \If{\rand{} $< \exp\left(\frac{\text{\cost - \neighbourCost}}{\text{\temp}}\right)$ }{
            \cost $\leftarrow$ \neighbourCost \;
            \mapping $\leftarrow$ \neighbour \;
        }
    }
    \temp $\leftarrow$ \temp $\times$ \Tconstant \;
}
\Return \finalMapping \;
}

\end{algorithm}

A modified version of simulated annealing has already been applied in \cite{zhou2020quantum} where a repetition parameter $R$ is used to explore several neighbours at each temperature step. The authors consider a simple \texttt{get\_neighbour} function that modifies randomly the current mapping $\pi$ to a neighbouring mapping $\pi_\text{neighbour}$. However, \texttt{get\_neighbour} function is limited as it is not aware of the underlying hardware. This means that from the set of mappings generated by this function and evaluated by the simulated annealing procedure, several mappings can be excluded even before evaluating the mapping cost.

We aim to improve the initial mapping generated with the simulated annealing procedure by designing a Hardware-aware Simulated Annealing (HSA) algorithm using a hardware-aware \texttt{get\_neighbour} method to explore the neighbouring mappings. 
To explore different mappings, we separate the \texttt{get\_neighbour} procedure in three algorithms governed by a top execution policy. This top layer policy decides which one of the three algorithms the \texttt{get\_neighbour} method should execute to obtain a new mapping. The policy we used randomly chooses which algorithm to use from the value of a random number.

The first algorithm, called \texttt{shuffle}, does not change the physical qubits involved in the current mapping but changes how they are mapped to logical qubits. The most straightforward algorithm that can be used for this task is a random shuffle -- we list the physical qubits involved in the mapping, randomly shuffle them, and obtain a new arbitrary mapping with the same physical qubits.

The second algorithm, \texttt{expand}, does not change the mapping between physical qubits and logical ones but replaces one of the physical qubits involved in the mapping by another physical qubit that is not part of the mapping. Instead of a hardware-unaware \texttt{expand}, we use an \texttt{expand} algorithm that tries to avoid separating the physical qubits in the current mapping into two disconnected groups. Moreover, the algorithm encourages re-arrangement of qubits based on the figure of merit chosen (i.e. final state fidelity, circuit depth, execution time). In this algorithm, we consider that strongly connected qubits have high fidelity. The hardware-aware implementation aims to identify the qubits with the least and most connections. Moreover, based on our tests with qubit measurement operations, we find there is a huge source of errors. To account for theses errors, we add weights to qubit to determine the best and worst qubits in terms of their measured fidelity.

The third algorithm used in the \texttt{get\_neighbour} algorithm is called \texttt{reset}. Its purpose is to give the possibility to the simulated annealing algorithm to escape local-minimums. This algorithm is needed because the first two algorithms \texttt{shuffle} and \texttt{expand} will likely explore only the close neighbourhood of the current mapping and may not be able to escape a local minimum. To avoid being stuck, the \texttt{reset} algorithm tries to find a potentially good new initial mapping from a randomly chosen qubit, without considering the previously explored mappings. The algorithm starts with a random qubit and expands the mapping by iteratively weighting all the qubits and adding the best qubit to the new mapping.

\subsection{Metrics}\label{sec:metrics}
To evaluate our solution and compare it to other algorithms, we use some metrics that are described in the following paragraphs.

The first metric is the success rate of the mapped quantum circuit on a given hardware. We define the success rate of a quantum circuit as the fidelity of the quantum state obtained at the end of the execution of this quantum circuit on the hardware. We estimate this success rate by executing the quantum circuit a large number of times ($8192$), counting the number of executions that gave the expected answer and dividing this number by the total number of executions. The expected answer is obtained by executing the quantum circuit on a simulator.

The second metric chosen is the additional number of \texttt{CNOT} gates. This metric is tightly linked with the total number of \texttt{SWAP} gates inserted.  

The third metric is the total execution time of the circuit. As the execution time of each \texttt{CNOT} gate can be extracted from~\cite{ibmq.hardwareinfos}, we can estimate the overall execution time for a given circuit. This metric is important for several reasons. First, it shows the ability of the mapping algorithm to schedule gates in parallel when possible and how good is the algorithm at doing this. Secondly, it allows us to have an idea of the importance of decoherence noise in the computed fidelity. For each qubit, the execution time is computed by adding the total execution time of gate operations acting on it. The longest qubit execution time is selected to represent the total execution time of the quantum circuit.

\begin{figure}
\centering
\includegraphics[width=0.25\textwidth]{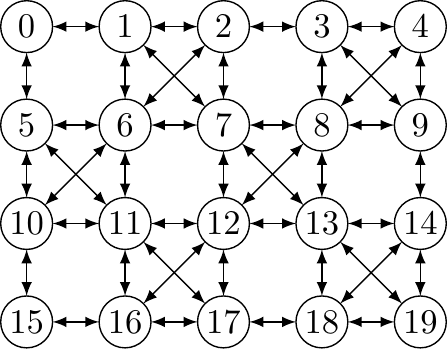}
\caption{ibmq\_tokyo topology.}
\label{fig:ibmq_tokyo}
\end{figure}

\section{Evaluation and comparison of the proposed HA Algorithm}
\label{section4}
\begin{figure*}
    \centering
    \subfloat[Number of additional gates\label{fig:gate_num_valencia}]{\scalebox{.75}{\includegraphics{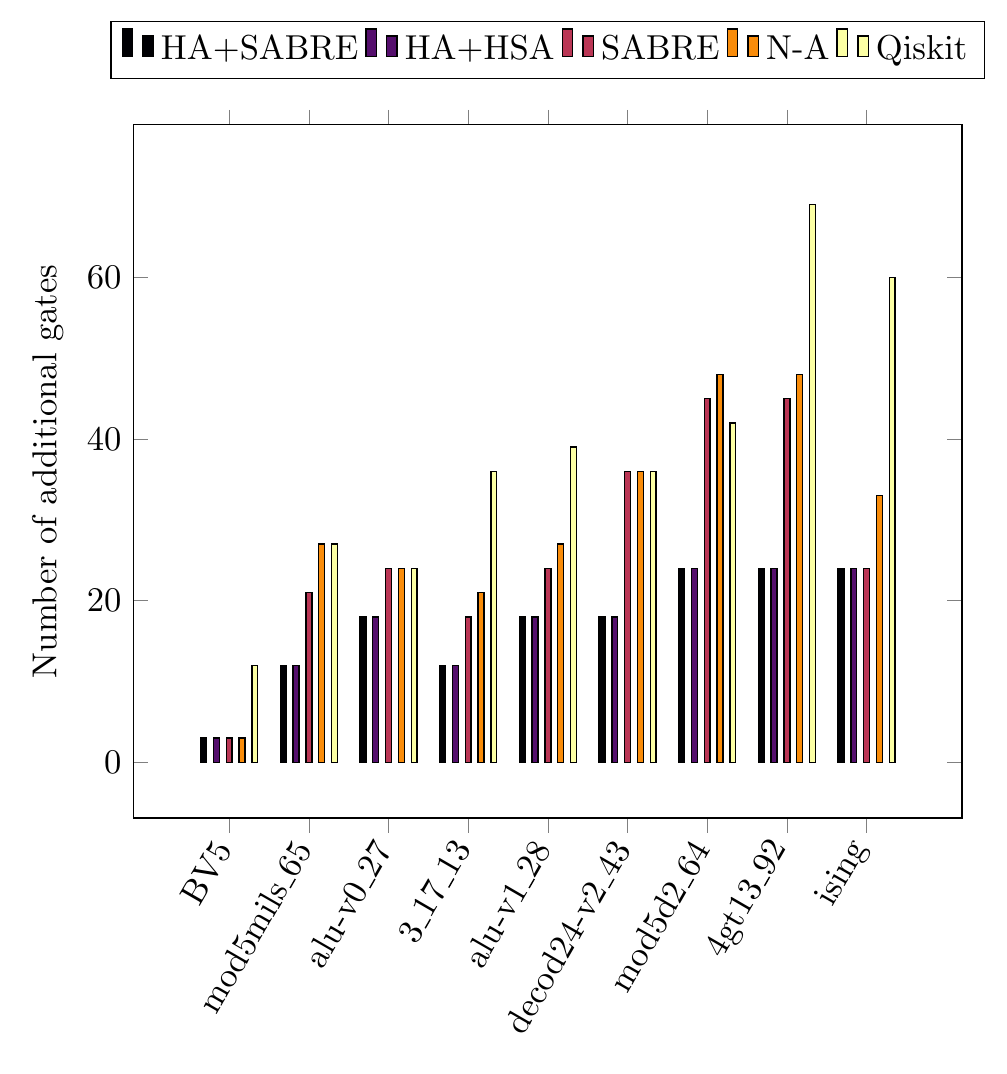}}}\hspace{0.1\textwidth}
    \subfloat[Fidelity\label{fig:fidelity_valencia}]{\scalebox{.75}{\includegraphics{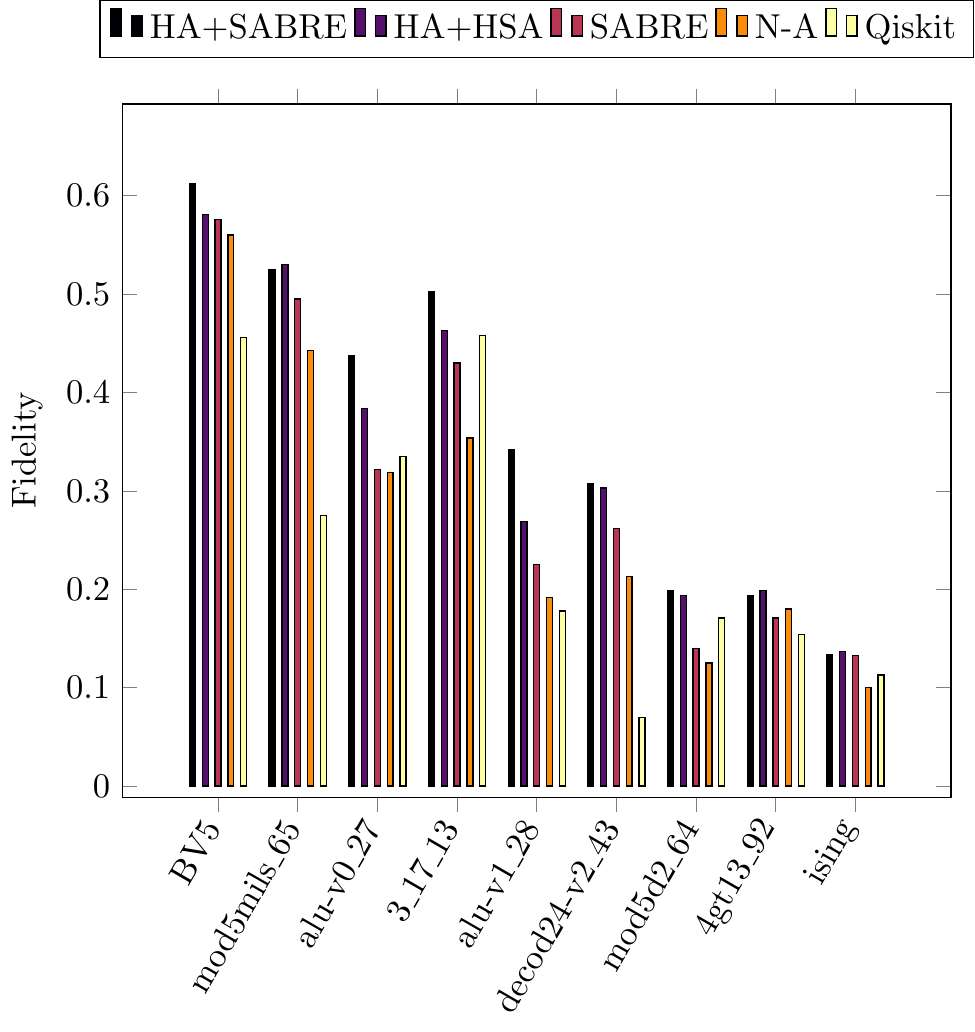}}}\qquad
    \caption{Comparison of number of additional gates and fidelity on ibmq\_valencia. HA has been used with $\alpha_1 = 0.5$, $\alpha_2 = 0.5$ and $\alpha_3 = 0$.}
    \label{fig:comparison_valencia}
\end{figure*}

\begin{figure*}
    \centering
    \subfloat[Number of additional gates\label{fig:gate_num_almaden}]{\scalebox{.75}{\includegraphics{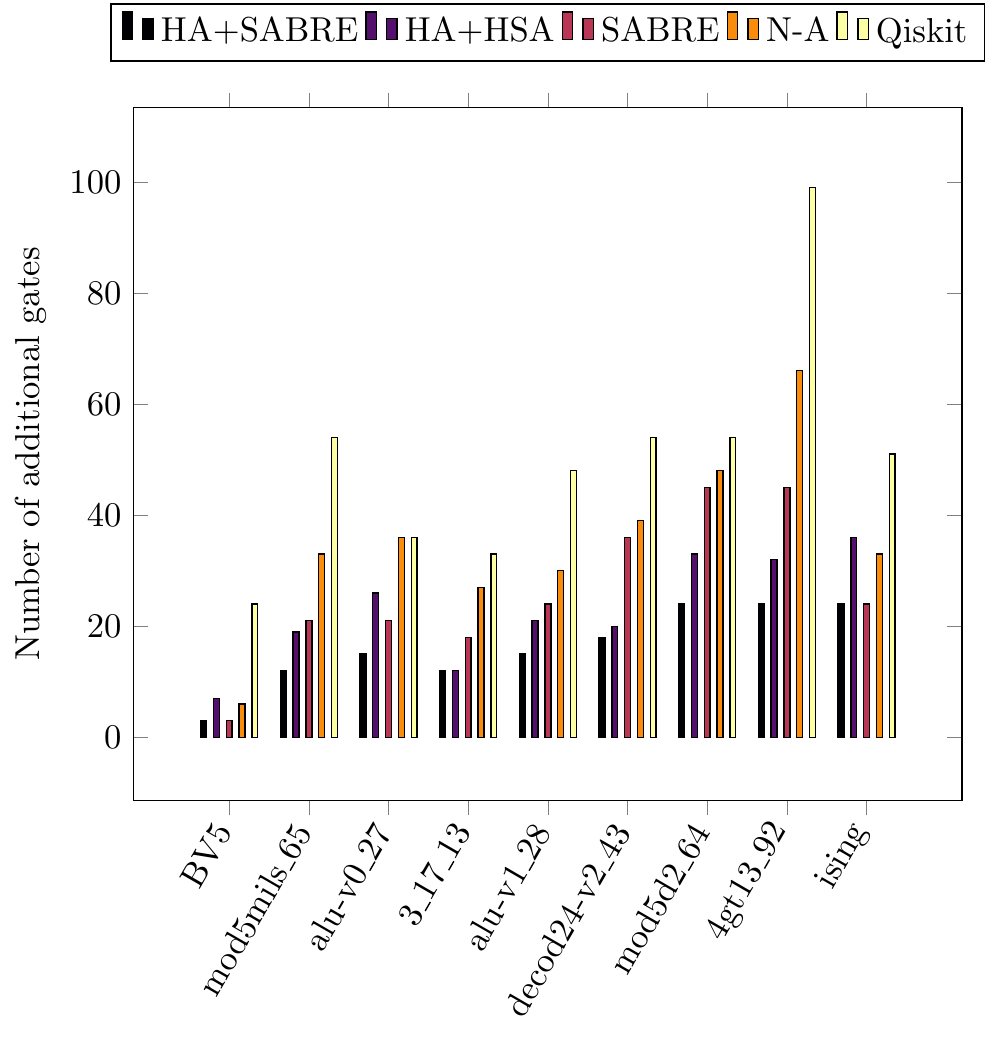}}}\hspace{0.1\textwidth}
    \subfloat[Fidelity\label{fig:fidelity_almaden}]{\scalebox{.75}{\includegraphics{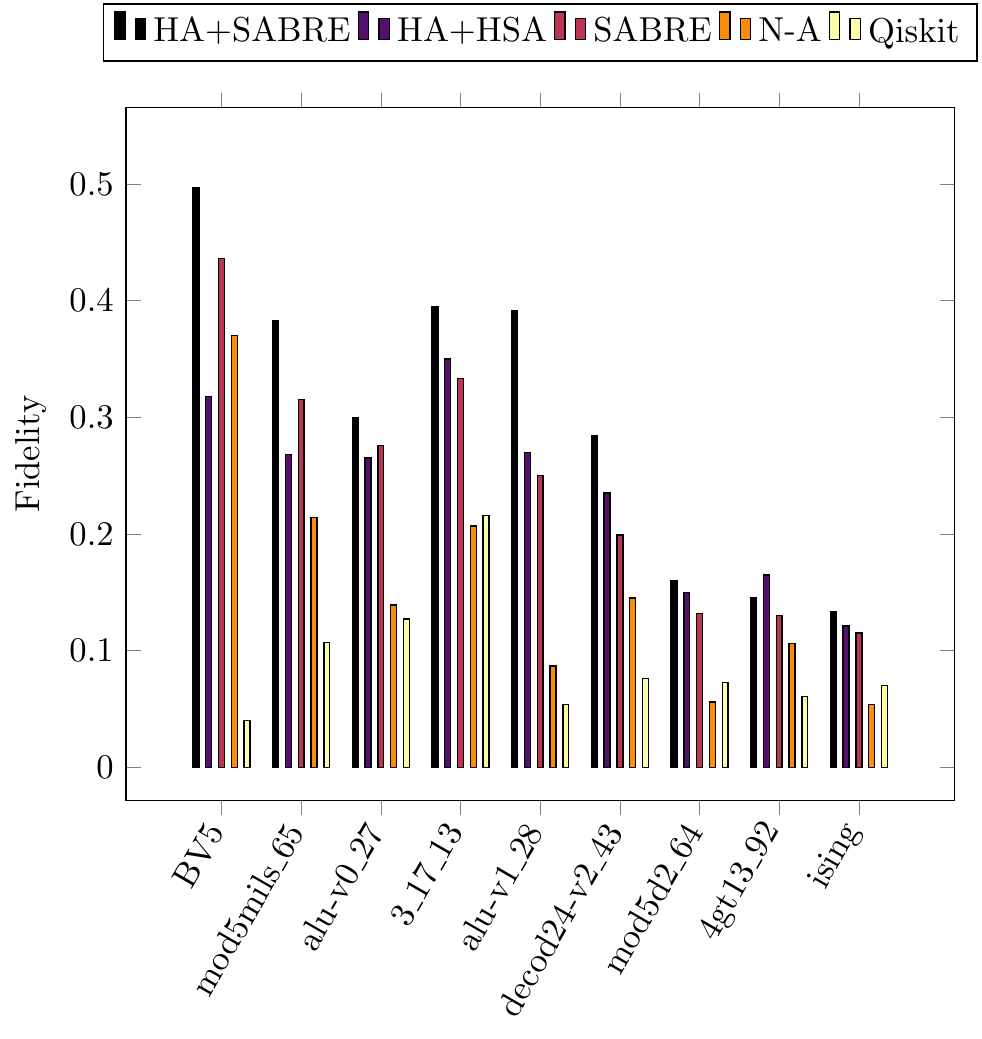}}}\qquad
    \caption{Comparison of number of additional gates and fidelity on ibmq\_almaden. HA has been used with $\alpha_1 = 0.5$, $\alpha_2 = 0.5$ and $\alpha_3 = 0$.}
    \label{fig:comparison_almaden}
\end{figure*}

\subsection{Methodology}
All the benchmarks used are collected from the previous works~\cite{zulehner2018efficient,li2019tackling,li2020qasmbench}. They include several functions taken from RevLib~\cite{WGT+:2008} as well as quantum algorithms from a variety of domains including optimization, simulation, quantum arithmetic, etc. They are well known in the community and given as quantum circuits written in the OpenQASM language~\cite{cross2017open}. 

We chose two quantum chips, ibmq\_almaden and ibmq\_valencia, available from IBM Quantum experience website and one quantum chip ibmq\_tokyo, which is not accessible currently but widely used by state-of-art algorithms. ibmq\_almaden is a 20-qubit quantum chip. Its topology and characteristics are summarised in Fig.~\ref{fig:ibmq20} and \autoref{tab:almaden_caracteristics}. ibmq\_valencia is a 5-qubit chip depicted in Fig.\autoref{fig:ibmq_valencia}. ibmq\_tokyo is a 20-qubit virtual chip depicted in Fig.~\ref{fig:ibmq_tokyo}. We execute benchmarks on ibmq\_valencia and ibmq\_almaden to check cases when the mapped quantum circuit needs all the available qubits or only a small number of them. Moreover, we use ibmq\_almaden and ibmq\_tokyo to compare our algorithm with state-of-art algorithms in terms of number of additional gates. Note that we do not have to execute the mapped quantum circuit on real quantum hardware to count the number of additional gates.

Our algorithm is implemented in Python and the Qiskit version is 0.19.1. To empirically evaluate our algorithm, we use a personal computer with 1 Intel i5-5300U CPU and 8 GB memory. The Operating System is Ubuntu 18.04.

Several published qubit mapping algorithms are available as discussed in \autoref{sec:introduction}. SABRE~\cite{li2019tackling} seems to be the best algorithm at the time of writing when comparing the number of inserted gates to make the quantum circuit hardware-compliant. It provides a good initial mapping method and a mapping transition algorithm. Another algorithm DL~\cite{dynamic-lookahead} (Dynamic look-ahead) based on SABRE shows an improvement in terms of number of additional gates. Moreover, the mapping method presented in~\cite{10.1145/3297858.3304075} uses the hardware calibration data to try to find a good mapping. We compare to all these algorithms. The source code of SABRE has been provided by the authors of the algorithm, and the mapping method presented in~\cite{10.1145/3297858.3304075}, called Noise-Adaptive (N-A) Compiler, has been integrated into Qiskit as a transpiler pass. Finally, we also include the default transpiler included in Qiskit as the baseline. We execute our HA mapping transition algorithm with two different initial mapping algorithm -- SABRE initial mapping algorithm and our Hardware-aware Simulated Annealing (HSA) algorithm. 

Summarising, to test on real hardware, five different algorithms are included in the benchmarks: 1) our HA mapping algorithm with SABRE initial mapping, 2) our HA mapping algorithm with HSA initial mapping, 3) SABRE mapping algorithm with SABRE initial mapping, 4) N-A Compiler and 5) Qiskit transpiler. For a fair comparison, we set the \texttt{optimisation\_level} parameter of the Qiskit transpiler to zero and make sure that the circuits obtained from the five methods are all executed with the same calibration data. The \texttt{optimisation\_level} is set to zero to invoke only the mapping transformation and not the optimisation transformation. Moreover, when using the N-A Compiler, the routing method is set to "lookahead" to make sure that it has the look-ahead ability. To compare the number of additional gates without accessing to real hardware, three algorithms are included: 1) SABRE, 2) DL, 3) HA.

To evaluate our algorithm with the different initial mapping methods, we allow each of them to call the mapping algorithm at most $100$ times. The number of calls to the mapping algorithm is a natural parameter of the simulated annealing-based method, but the SABRE initial mapping method only needs two calls. To let the SABRE algorithm take advantage of a larger number of calls, we repeat the algorithm on several random initial mappings until no more calls are allowed and choose the best mapping found. The whole process is repeated $10$ times to obtain $10$ initial mappings. 

We divide benchmarks by size according to their number of gates. We only execute small size benchmarks on real quantum hardware, because the other benchmarks with a large number of gate operations introduce too much noise to obtain any meaningful results. Moreover, the initial mapping generation process described above is applied on small and medium sized benchmarks. Large benchmarks suffer from long run time, so we generate $10$ initial random mappings and use them with different algorithms. When using ibmq\_tokyo virtual chip, we select the best results out of five attempts which is a similar approach applied in SABRE and DL.

When testing the HSA algorithm we used a random policy to choose which one of the three subroutines to execute. The \texttt{shuffle} procedure is executed with a probability of $0.9$, the \texttt{expand} algorithm is chosen with a probability $0.08$ and the \texttt{reset} procedure is executed when the two previous algorithms are not used (i.e. $0.02$ probability).

First, we compare the number of additional gates and fidelity. The weight parameter $\alpha_1$ of swap matrix $S$ is set to 0.5, the weight parameter $\alpha_2$ of CNOT error matrix $\mathcal{E}$ is set to 0.5 and the weight parameter $\alpha_3$ of CNOT execution time $T$ is set to 0. Second, we compare the number of additional gates and total execution time. Weight parameter $\alpha_1$ of swap matrix $S$ is set to 0.5, weight parameter $\alpha_2$ of CNOT error matrix $\mathcal{E}$ is set to 0 and weight parameter $\alpha_3$ of CNOT execution time $T$ is set to 0.5. Third, we compare the number of additional gates for circuits that are not executable on the real quantum device. Weight parameter $\alpha_1$ of swap matrix $S$ is set to 1, and the other two parameters are set to 0. For these three comparisons, the weight parameter $W$ in the cost function is set to 0.5 and size of extended layer is set to 20.

\subsection{Experimental results}

We compare both the average number of additional gates (see Fig.\autoref{fig:gate_num_valencia} and Fig.\autoref{fig:gate_num_almaden}) and average output state fidelity (see Fig.\autoref{fig:fidelity_valencia} and Fig.\autoref{fig:fidelity_almaden}) among the 10 initial mappings for the five methods. The complete experimental results are listed in ~\autoref{tab:result_valencia} and~\autoref{tab:result_almaden}.

The Qiskit default qubit mapping algorithm is nearly always the worst one in terms of additional gates, which translates in most of the cases to the worst output state fidelity. Although N-A Compiler takes into account the calibration data and has the look-ahead ability, results show that it does not outperform the SABRE mapping algorithm with SABRE initial mapping (labelled as SABRE in the plots). Our HA mapping algorithm with SABRE initial mapping (labelled as HA+SABRE in the plots) seems to be the best combination as in average it achieves the best output state fidelity. Moreover, our HA algorithm with SABRE initial mapping gives the minimum number of additional gates. HA mapping algorithm with HSA initial mapping (labelled as HA+HSA in the plots) is also good, but its results are less consistent than HA+SABRE due to its random nature. Although, in many test cases, it outperforms SABRE.

We also tried to map and execute the \texttt{qft\_10} circuit. We found that its output fidelity is less than $0.01$ for all the methods tested in the benchmark. Because the base fidelity is too low to perform a meaningful comparison, we only compare the number of additional gates as summarised in~\autoref{tab:additional_gates} and~\autoref{tab:additional_gates2} for quantum circuits with a medium-to-large number of gates.

\begin{figure*}
    \centering
    \subfloat[Execution time\label{fig:execution_time_valencia}]{\scalebox{.5}{\includegraphics{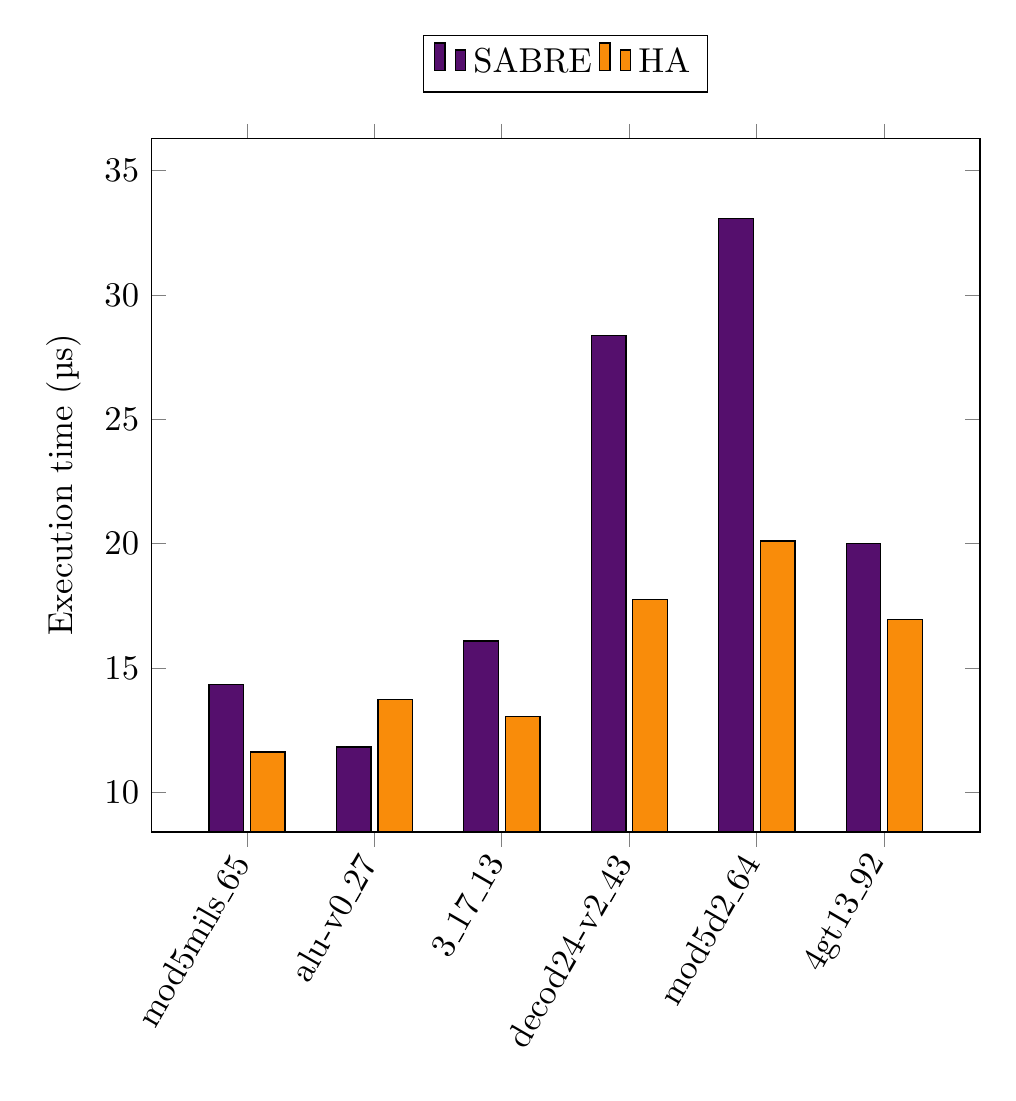}}}\qquad
    \subfloat[Number of additional gates\label{fig:execut_gate_num_valencia}]{\scalebox{.5}{\includegraphics{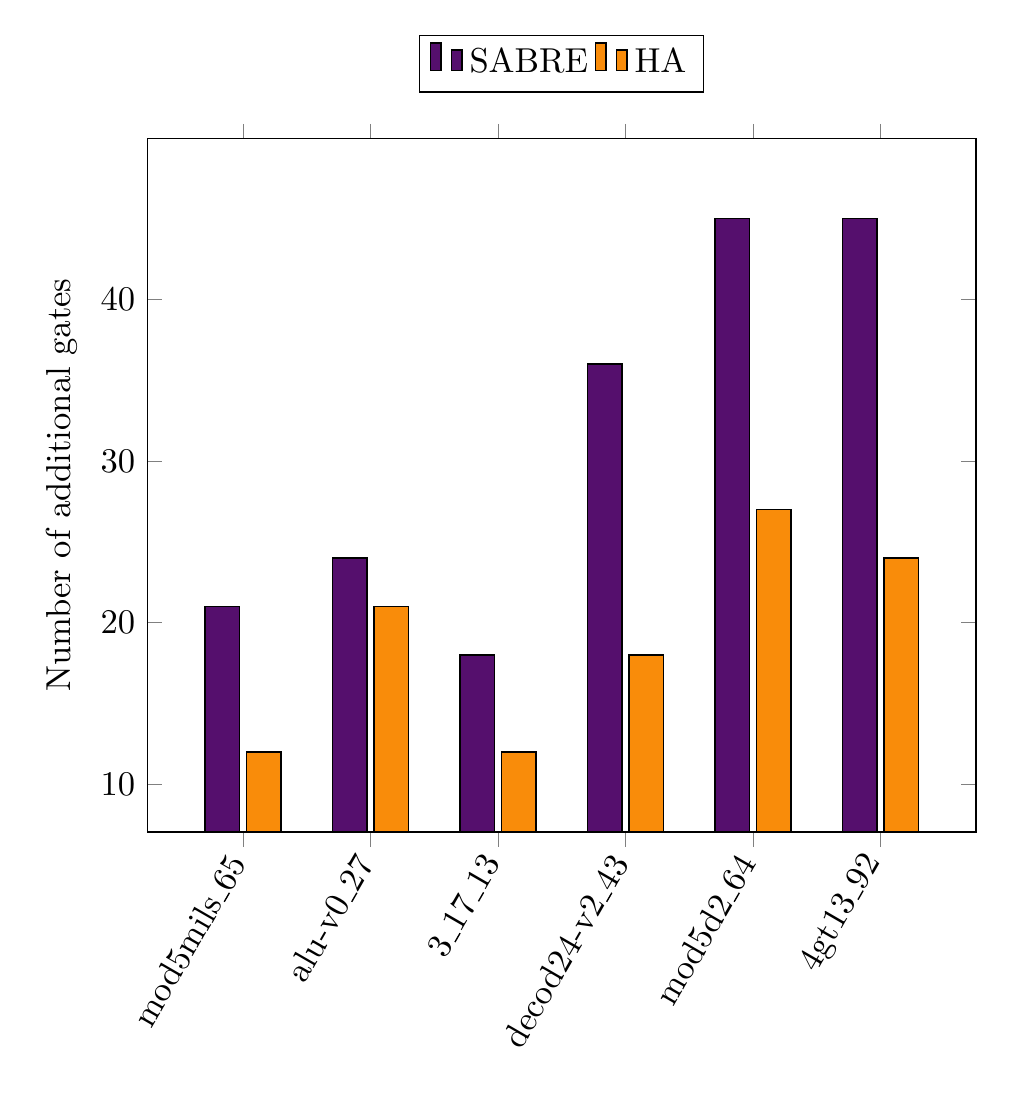}}}\qquad
    \subfloat[Fidelity\label{fig:execut_fidelity_valencia}]{\scalebox{.5}{\includegraphics{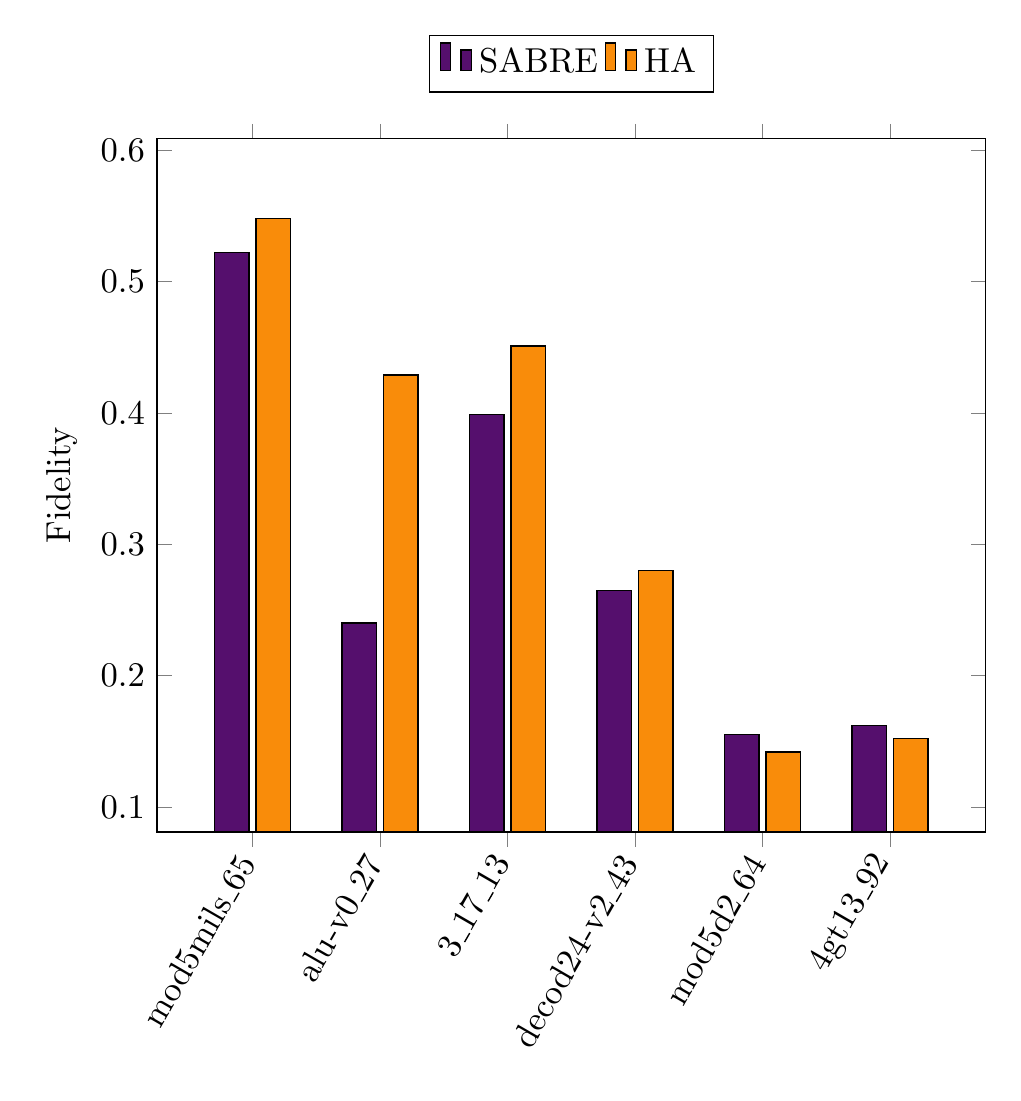}}}\qquad
    \caption{Comparison of execution time, number of additional gates and fidelity on ibmq\_valencia. HA has been used with $\alpha_1 = 0.5$, $\alpha_2 = 0$ and $\alpha_3 = 0.5$.}\label{fig:comparison_execution_time}
\end{figure*}

Fig.~\ref{fig:comparison_execution_time} shows the result of comparing the execution times, number of additional gates and fidelities of our HA algorithm with SABRE algorithm on ibmq\_valencia. The execution time is reduced by 19\% on average. Even though the weight parameter $\alpha_2$ of CNOT error matrix $\mathcal{E}$ is set to 0, the fidelity is improved by 8\%. The number of additional gates is reduced by 38\%.
    
\autoref{tab:additional_gates} lists the result of the number of additional gates on ibmq\_almaden. Using the selection of \texttt{SWAP} and \texttt{Bridge} gate, our HA algorithm can outperform SABRE on circuits with different sizes. For medium circuits, HA gives similar results as SABRE and an improvement from SABRE for only one circuit among the eight circuits tested. For large circuits, HA outperforms SABRE and consistently reduces the number of additional gates by $28\%$ on average. \autoref{tab:additional_gates2} shows the number of additional gates on ibmq\_tokyo when comparing our HA algorithm with SABRE and DL. DL outperforms SABRE and our HA algorithm can further reduce the number of additional gates by $14\%$ on average. 
SABRE and DL only provide their runtime on ibmq\_tokyo, the difference between runtime of the three algorithms is shown in \autoref{tab:additional_gates2}. Note that, DL is written in C++ and tested on a normal personal computer. SABRE is written in Python and tested on a server with 2 Intel Xeon E5-2680 CPUs (48 logical cores) and 378GB memory. Since there is an intrinsic speed difference between C++ and Python as well as the different devices used, the runtime data in this table are for reference rather than for comparison.

\begin {table*}
\caption {Number of additional gates on ibmq\_almaden for large circuits. HA has been used with $\alpha_1 = 1$, $\alpha_2 = 0$ and $\alpha_3 = 0$.} \label{tab:additional_gates} 
\begin{center}
\resizebox{\textwidth}{!}{%
\begin{threeparttable}
\centering
\resizebox{\textwidth}{!}{%
\includegraphics{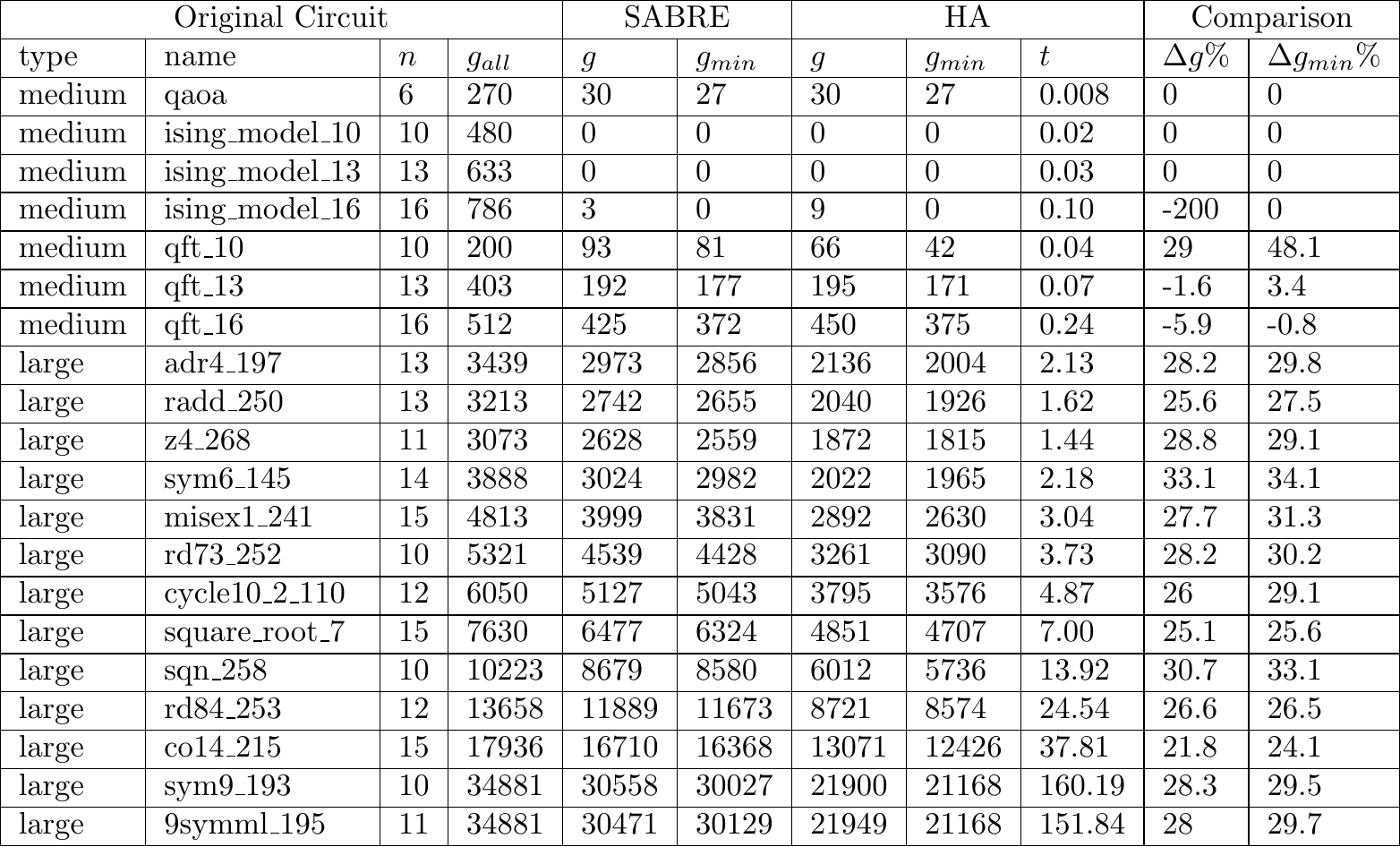}
}
\begin{tablenotes}
    \footnotesize
    \item $\mathbf{n}$: number of qubits. $\mathbf{g_{all}}$: total number of gates. $\mathbf{g}$: average number of additional gates. $\mathbf{g_{min}}$: minimum number of additional gates. $\mathbf{t}$: runtime in seconds. $\mathbf{\Delta{g}}$: comparison of average number of additional gates between HA and SABRE. $\mathbf{\Delta{g_{min}}}$: comparison of minimum number of additional gates between HA and SABRE.
\end{tablenotes}
\end{threeparttable}
}
\end{center}
\end{table*}

\begin {table*}
\caption {Number of additional gates on ibmq\_tokyo for large circuits. HA has been used with $\alpha_1 = 1$, $\alpha_2 = 0$ and $\alpha_3 = 0$.} \label{tab:additional_gates2} 
\begin{center}
\resizebox{\textwidth}{!}{%
\begin{threeparttable}
\centering
\resizebox{\textwidth}{!}{%
\includegraphics{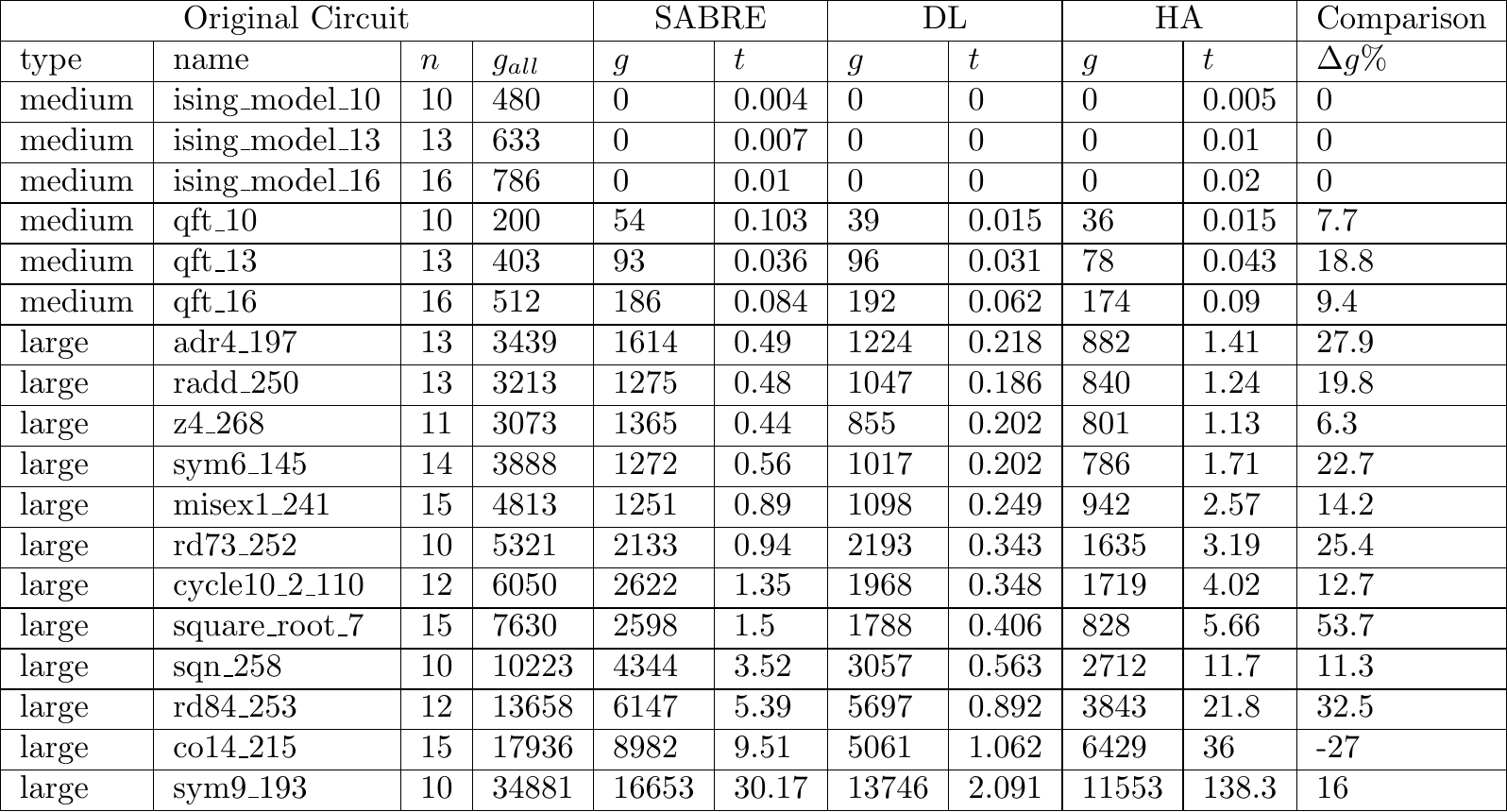}
}
\begin{tablenotes}
    \footnotesize
    \item $\mathbf{n}$: number of qubits. $\mathbf{g_{all}}$: total number of gates. $\mathbf{g}$: minimum number of additional gates. $\mathbf{t}$: runtime in seconds. $\mathbf{\Delta{g}}$: comparison of minimum number of additional gates between HA and DL.
\end{tablenotes}
\end{threeparttable}
}
\end{center}
\end{table*}

\begin{table*}[ht]
\caption {Comparison of number of additional gates and fidelity on ibmq\_valencia. HA has been used with $\alpha_1 = 0.5$, $\alpha_2 = 0.5$ and $\alpha_3 = 0$.}
\label{tab:result_valencia}
\begin{center}
\resizebox{\textwidth}{!}{%
\begin{threeparttable}
\centering
\resizebox{\textwidth}{!}{%
\includegraphics{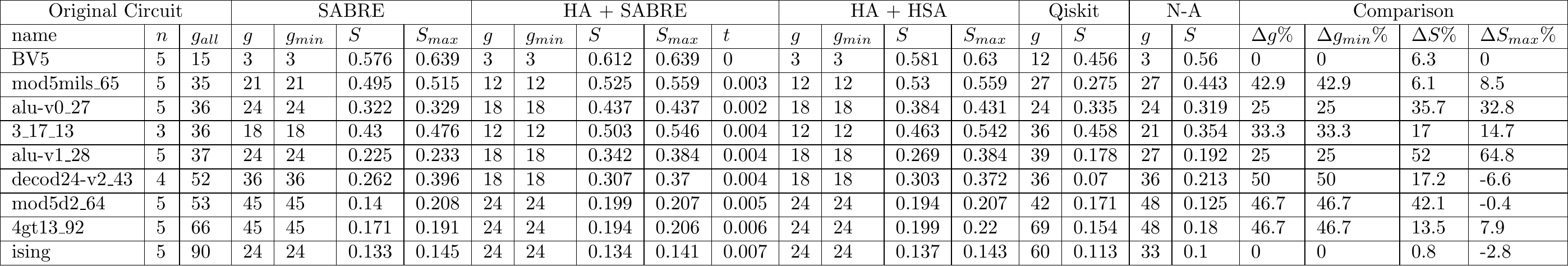}
}
\begin{tablenotes}
      \footnotesize
      \item $\mathbf{n}$: number of qubits. $\mathbf{g_{all}}$: total number of gates. $\mathbf{g}$: average number of additional gates. $\mathbf{g_{min}}$: minimum number of additional gates. $\mathbf{S}$: mean of success rate. $\mathbf{S_{max}}$: maximum of success rate. $\mathbf{\Delta{g}}$: comparison of average number of additional gates between HA+SABRE and SABRE. $\mathbf{\Delta{g_{min}}}$: comparison of minimum number of additional gates between HA+SABRE and SABRE. $\mathbf{\Delta{S}}$: comparison of mean of success rate between HA+SABRE and SABRE. $\mathbf{\Delta{S_{max}}}$: comparison of maximum of success rate between HA+SABRE and SABRE.
      $\mathbf{t}$: runtime of HA+SABRE in seconds.
\end{tablenotes}
\end{threeparttable}
}
\end{center}
\end{table*}

\begin{table*}[ht]
\caption{Comparison of number of additional gates and fidelity on ibmq\_almaden. HA has been used with $\alpha_1 = 0.5$, $\alpha_2 = 0.5$ and $\alpha_3 = 0$.}
\label{tab:result_almaden}
\begin{center}
\resizebox{\textwidth}{!}{%
\begin{threeparttable}
\centering
\resizebox{\textwidth}{!}{%
\includegraphics{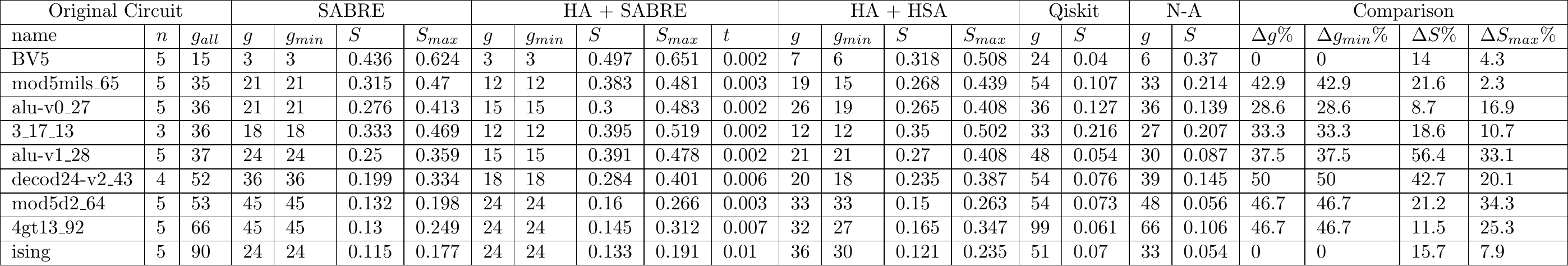}
}
\begin{tablenotes}
    \footnotesize
      \item $\mathbf{n}$: number of qubits. $\mathbf{g_{all}}$: total number of gates. $\mathbf{g}$: average number of additional gates. $\mathbf{g_{min}}$: minimum number of additional gates. $\mathbf{S}$: mean of success rate. $\mathbf{S_{max}}$: maximum of success rate. $\mathbf{\Delta{g}}$: comparison of average number of additional gates between HA+SABRE and SABRE. $\mathbf{\Delta{g_{min}}}$: comparison of minimum number of additional gates between HA+SABRE and SABRE. $\mathbf{\Delta{S}}$: comparison of mean of success rate between HA+SABRE and SABRE. $\mathbf{\Delta{S_{max}}}$: comparison of maximum of success rate between HA+SABRE and SABRE.
      $\mathbf{t}$: runtime of HA+SABRE in seconds.
\end{tablenotes}
\end{threeparttable}
}
\end{center}
\end{table*}

\section{Discussion}
\subsection{Design guideline}
Given the current available NISQ hardware, it is important to adapt quantum programs to execute on such hardware while taking into account their physical constraints and limitations (noisy operations, number of qubits and gates). Here, we list several guidelines that can help a programmer to design quantum circuits that comply on given quantum hardware. \\
\begin{itemize}
\item Check the topology and the calibration data of the device targeted. Try to map the most used qubit of the mapped circuit to the physical qubit that has the strongest coupling connection.\\
\item Try to apply a \texttt{CNOT} gate on qubits that are directly connected and with a reliable (i.e. low error rate) interconnect, so that no more additional gates are needed, and the overall circuit fidelity is improved. \\
\item If a \texttt{CNOT} gate cannot be applied on two-adjacent qubits, try to apply on two qubits whose distance is two on the coupling graph. In such situation, one can select between a \texttt{SWAP} and \texttt{Bridge} gate to execute the \texttt{CNOT} gate. Also, the number of additional gates will be reduced.
\end{itemize}

\subsection{Future work}
In this work, we present an efficient hardware-aware mapping algorithm based on heuristic search. For future studies, we find that the following potential research directions can be explored.
First, our HA algorithm only takes into consideration the calibration data, which includes the gate error and the execution time. However, other physical constraints, such as crosstalk error may be included to take into account crosstalk coupling between interconnects. 
Secondly we would like to investigate the adaptation of such a mapping algorithm to a multi-programming mechanism as introduced in~\cite{das2019case}. Executing multiple quantum circuits on the same chip allows us to use more efficiently hardware resources but may decrease the fidelity of the quantum operations due to unwanted interactions. 
Finally, we find it relevant to investigate mapping algorithms for specific use cases such as quantum circuits constructed for quantum chemistry computations with VQE~\cite{vqe-original} or to solve linear systems with the VQLS algorithm~\cite{vqls-bravo-prieto}.

\section{Conclusion}
The quantum computers are now in the NISQ era. There's a gap between the design and execution of a quantum circuit in NISQ hardware. In this paper, we present a hardware-aware heuristic for qubit mapping problem that adapts the quantum circuit to the quantum hardware. We design a mapping transition algorithm that uses calibration data and selects from either a \texttt{SWAP} or \texttt{Bridge} gate for qubit movement. Experimental results show that our algorithm can outperform state-of-the-art algorithms in terms of the number of additional gates, fidelity and execution time. Our algorithm is evaluated on IBM quantum devices, but should be general enough to be used on quantum devices from other vendors as well.

\section*{Acknowledgment}
This work is funded by the QuantUM Initiative of the Region Occitanie, University of Montpellier and IBM Montpellier as well as by a research collaboration grant between TOTAL, LIRMM and CERFACS. We would like to thank the authors of SABRE for the meaningful discussions and exchanges.

\section*{Supplementary Information}\label{supplementary}
Authors have made available the source code and it can be found at the following link: \url{https://github.com/peachnuts/HA}.
\bibliography{bibliography}{}
\bibliographystyle{plain}

\end{document}